\documentclass[twocolumn]{aastex631}

\newcommand{\caii}{Ca\,{\sc ii}}
\newcommand{\mgii}{Mg\,{\sc ii}}

\begin{document}

%\title{Swift X-Ray and UV Observations of Proxima Centauri's Stellar Cycle}
\title{X-Ray, UV, and Optical Observations of Proxima Centauri's Stellar Cycle}

\correspondingauthor{Brad Wargelin}
\email{bwargelin@cfa.harvard.edu}

\author[0000-0002-2096-9586]{Bradford J.~Wargelin}
%\author{Bradford J.~Wargelin}
\affiliation{Center for Astrophysics|Harvard \& Smithsonian, 
60 Garden Street, Cambridge, MA 02138, USA}

\author[0000-0001-7032-8480]{Steven H.~Saar}
%\author{Steven H.~Saar}
\affiliation{Center for Astrophysics|Harvard \& Smithsonian, 
60 Garden Street, Cambridge, MA 02138, USA}

\author{Zackery A.~Irving}
\affiliation{School of Physics and Astronomy, University of Southampton,
University Road, Southampton SO17 1BJ, UK}

\author[0000-0002-7597-6935]{Jonathan D.~Slavin}
%\author{Jonathan D.~Slavin}
\affiliation{Center for Astrophysics|Harvard \& Smithsonian, 
60 Garden Street, Cambridge, MA 02138, USA}

\author{Peter Ratzlaff}
\affiliation{Center for Astrophysics|Harvard \& Smithsonian, 
60 Garden Street, Cambridge, MA 02138, USA}

\author[0000-0001-7804-2145]{Jos\'{e}-Dias do Nascimento, Jr}
%\author{Jos\'{e}-Dias do Nascimento, Jr}
\affiliation{Center for Astrophysics|Harvard \& Smithsonian, 
60 Garden Street, Cambridge, MA 02138, USA}
\affiliation{Univ.~Federal do Rio G.~do Norte, UFRN, Dep.~de F\'{i}sica,
CP 1641, 59072-970, Natal, RN, Brazil}



\begin{abstract}

Proxima Cen (GJ 551; dM5.5e) is one of only about a dozen fully convective 
stars known to have a stellar cycle, and the only one to have long-term X-ray 
monitoring. A previous analysis found that X-ray and mid-UV observations, 
particularly two epochs of data from Swift, were consistent with a
well sampled $\sim$7 yr optical cycle seen in ASAS data, 
but not convincing by themselves. 
The present work incorporates several years of new ASAS-SN optical data and
an additional five years of Swift XRT and UVOT observations, 
with Swift observations now spanning 2009 to 2021 and optical
coverage from late 2000.  X-ray observations by XMM-Newton
and Chandra are also included.
Analysis of the combined data, which includes modeling and
adjustments for stellar contamination in the optical and UV,
now reveals clear cyclic behavior in
all three wavebands with a period of 8.0 yr.
We also show that UV and X-ray intensities are anti-correlated with optical
brightness variations caused by the cycle and by rotational modulation, 
discuss possible indications of two coronal mass ejections,
and provide updated results for the previous finding of a 
simple correlation between X-ray cycle amplitude and Rossby number over
a wide range of stellar types and ages.

\end{abstract}

\section{Introduction} \label{sec:Intro}
%%%%%%%%%%%%%%%%%%%%%%%%%%%%%%%%%%%%%%%%%%%%%%%%%%%%%%%%%%%%%%%

Despite accounting for $\sim$70\% of the stellar population, 
M stars have until very recently been poorly represented in studies
of magnetic activity cycles because of their intrinsic faintness.
In the pioneering HK Project at Mount Wilson Observatory \citep{Baliunas1995},
which began in 1966 and monitored 
chromospheric \caii\ H and K lines (3969 and 3934 \AA) in roughly 300 stars,
only one of the objects of study was an M star (Lalande 21185; dM2).
Technological progress, however, has steadily brought an increasing
number of M stars under scrutiny in spectroscopic
programs such as HARPS \citep{Mayor2003}, 
%McDonald Observatory (MDO) M Dwarf Planet Search (Cochran \& Hatzes 1993)
MDO Planetary Search \citep{Cochran1993}, and 
CASLEO HKalpha \citep{Cincunegui2004},
and in photometric monitoring programs such as
the All Sky Automated Survey project
\citep[ASAS;][]{Pojmanski1997,Pojmanski2002},
ASAS for SuperNovae
\citep[ASAS-SN;][]{Shappee2014,Kochanek2017},
%and Asteroid Terrestrial-impact Last Alert System
%\citep[ATLAS;][]{Tonry2018}.
and ATLAS \citep{Tonry2018}.
Although discovery of planets via radial velocity measurements
or detection of transient behavior such as supernovae are generally
the focus of such projects, their sustained measurements over many
years often lend themselves to studies of cyclic behavior as well.

Using roughly a decade of ASAS data, \citet{SM2016} reported
on apparent cycles in around 40 stars, half of which were M stars.
Of those, around a dozen were fully convective, with stellar
type M3.5 or later.
This was a surprising result, since
most theories of stellar magnetism
predict that cyclic behavior can only be supported by
solar-type $\alpha\Omega$ dynamos, which are driven by magnetic
shear at a radiative/convective boundary, or tachocline
\citep{Dikpati1999}.
Fully convective stars, of course, do not have tachoclines, 
and instead their magnetic fields are expected to be driven
by $\alpha^2$ dynamos.
Some theoretical work, however,
suggests that $\alpha^2$ dynamos can
in fact support activity cycles under certain conditions
\citep{Rudiger2003,Chabrier2006,Gastine2012,Kapyla2013,Yadav2016}, 
and observations show
that fully convective stars follow the same rotation-activity
relation as partially convective stars \citep{Wright2016}.
The presence of cyclic behavior in the $\sim$dozen fully convective stars
noted by \citet{SM2016} was 
%The observational results of \citet{SM2016} for fully convective stars were
recently confirmed by \cite{Irving2023}, who combined the 
original ASAS data with several years of later ASAS-SN data.
Three of those cycles plus
one other (GJ876; M3.5) were also reported by
\citet{Mignon2023} based on chromospheric emission lines
(H$\alpha$, \caii\ H\&K, or the Na D doublet).

A common target of those studies is Proxima Cen (dM5.5).
%One of the fully convective stars examined in those papers is
%Proxima Cen (dM5.5).  
\citet{SM2016} analyzed nine years of Proxima ASAS-3 data 
and found a cycle with a period of $6.8\pm0.3$ yr, 
and \citet{Wargelin2017} found a period of $7.05\pm0.15$ yr after including 
five additional years of ASAS-4 data. The latter paper
also analyzed data from several
X-ray missions, particularly two seasons of Swift observations
from 2009/2010 and 2012/2013, and reported that measured intensities were
consistent with a stellar cycle opposite in
phase to the optical cycle, with hints of the same anticorrelation for
rotational modulation.

As the star nearest the Sun at a distance of 1.302 pc,
Proxima is by far the most easily studied of late-type M's, and is
a subject of particular interest for its possession of
one confirmed planet \citep{AngladaEscude2016}
and two additional candidates \citep{Damasso2020,Faria2022}.
%\textcolor{red}{
%Proxima b is reported as having a mass of **TBD (**REF),
%and the less firmly established Proxima c with $m=TBD$ (**REF).
%Adopting an inclination of $47^{\circ}\pm7^{\circ}$ 
%reported by \citet{Klein2021},
%the planetary masses are respectively *** do the math**.}
Proxima is also notable as the only fully convective star
that has been monitored in X rays over at least one full
stellar cycle, making it unique among an already small set of stars 
to have their cycles measured at high energies,
where cycle amplitudes are much larger than in the optical
and more directly tied to magnetic activity.
%Although observing at X-ray and UV energies 
%(i.e., above the Earth's atmosphere) is a much bigger undertaking
%than in the optical, stellar X-ray/UV emission is driven by
%magnetic fields and therefore much more directly tied to magnetic
%activity and cycles than optical photometry, in which the
%relative dimness of star spots competes against the brightness
%of the umbrae surrounding them.  

%The vast majority of stellar cycle studies rely on optical
%data, either spectral or photometric, and fewer than ten
%stars have had X-ray monitoring.

In Section~\ref{sec:Data} we describe the assembly and analysis of
optical data from ASAS and ASAS-SN (\ref{subsec:DataOptical})
and X-ray (\ref{subsec:DataXray}) and UV data (\ref{subsec:DataUVOT})
from Swift and other missions,
followed by a study of brightness correlations among those
data sets on rotational time scales (\ref{subsec:RotationalMod}).
In Section~\ref{sec:Results} we measure cycle properties in all three
wavebands, compare the derived X-ray cycle amplitude with
those from other stars of varying rotational rates and stellar types,
and discuss possible signatures of coronal mass ejections,
followed by a summary in Section~\ref{sec:Summary}.
%\textcolor{red}{Add something about unique info provided by X-ray/UV
%cycle obs's once I've finished the end section(s).}

%%%%%%%%%%%%%%%%%%%%%%%%%%
%See other papers for stuff to discuss and cite.
%Mag field generation and cycles, obs monitoring (CaII, photom), Proxima 
%b and c planets and activity/atmo stripping,
%M stars, recent discovery of fully convective cycles (SM2016, Warg2017,
%Zac's paper), 
%higher cycle
%amplitudes in UV and Xray, few cases of such monitoring, previous
%work on Proxima (83-day rotation, etc.), 
%cases of X-ray cycles being more regular than CaII (and/or mention
%in Rotational Modulation/Correlation section), Section outline.
%Mention inclination measurement here if the end of 2.4.2 gets chopped.

%%%%%%%%%%%%%%%%%%%%%%%%%%%%%%%%%%%%%%%%%%%%%%%%%%%%%%%%%%%%%%%
\section{Observational Data and Analysis} \label{sec:Data}
%%%%%%%%%%%%%%%%%%%%%%%%%%%%%%%%%%%%%%%%%%%%%%%%%%%%%%%%%%%%%%%

%%%%%%%%%%%%%%%%%%%%%%%%%%%%%%%%%%%%%%%%%%%%%%%%%%%%%%%%%%%%%%%
\subsection{Optical Data} \label{subsec:DataOptical}
%%%%%%%%%%%%%%%%%%%%%%%%%%%%%%%%%%%%%%%%%%%%%%%%%%%%%%%%%%%%%%%

A previous study of Proxima's cycle \citep{Wargelin2017} 
used $V$-band optical data from ASAS,
%the All Sky Automated Survey project
%\citep[ASAS;][]{Pojmanski1997,Pojmanski2002},
specifically the publicly available ASAS-3 data 
covering Dec 2000 to Sep 2009, plus five years of 
ASAS-4 data (Jul 2010 to Aug 2015;
private communication, G.~Pojma\'{n}ski) which were cross-calibrated with
ASAS-3 using 33 nearby stars.  
For the present work we include
additional ASAS-4 data extending into 2019
that were also provided for \citet{Damasso2020},
but most of the new optical data come from the ASAS-SN program.
%the ASAS for SuperNovae project
%\citep[ASAS-SN;][]{Shappee2014,Kochanek2017}. 
ASAS-SN $V$-band observations 
of Proxima's field began in Mar 2016 and ran through Aug 2018, 
overlapping with a switch to the $g$ band beginning in Jun 2018.

$V$-band ASAS-3 data were downloaded from
{\url https://www.astrouw.edu.pl/asas/} and ASAS-4 data were
provided by the program leader, G.~Pojma\'{n}ski.
As recommended,
we used only measurements with A or B quality grades,
and chose brightness measurements using 
the 1\arcmin\ aperture (MAG\_2; 4 pixel diameter), 
which had the lowest scatter among the five aperture choices.
Typical ASAS resolution is 23\arcsec\ FWHM.

%Of the 1462 measurements with A or B quality flags we kept
%only those with magnitudes that fell within 3 standard deviations
%of the mean (grouped by observing season, which approximately
%coincides with calendar year).
%The remaining 1433 observations, typically 3 minutes long,
%were made on 1085 nights.

$V$- and $g$-band ASAS-SN data were downloaded from the ASAS-SN Sky Patrol site
({\url https://asas-sn.osu.edu/}) in multiple steps using 
coordinates accurate to within 1\arcsec\ over discrete time intervals
%proper-motion-corrected coordinates
%({\url https://asas-sn.osu.edu/}) in multiple steps using 
%proper-motion-corrected coordinates
(generally two intervals per observing season)
and then collated into a single file.
% with positions accurate to within 1\arcsec.
(ASAS extractions automatically account for proper motion, which is
3.86\arcsec per yr for Proxima.) ASAS-SN brightness extractions use a
% 3/6/24:  D'oh!  It's 16" radius.
32\arcsec-diameter aperture, approximately double the typical telescope PSF FWHM.

To remove outliers, mostly from flares, we clipped measurements more than
$2\sigma$ from the seasonal average in ASAS data. For ASAS-SN, 
% we followed 
%the procedure described by \citet{Irving2023}, first clipping measurements 
%more than $2 \sigma$ from the average over Proxima's 83 day rotation period, 
we clipped measurements more than $2.5 \sigma$ from the seasonal average,
and then more than $2.5 \sigma$ from the average in 7 day bins.
%{\it Zac, were these sliding boxcars for ASAS-SN?  -- nope, they were non-overlapping windows.}

The next step was to combine
the ASAS and ASAS-SN data, which were collected using different filters
and, for ASAS-SN, several telescopes around the world.  
The $V$ and $g$ bands are not far
apart (central wavelengths of 551 and 520 nm) and for a first approximation
we used the relations of 
\citet{Kent1985} and \citet{Windhorst1991},
\begin{equation}
    {\rm V} = {\rm g} - 0.03 - 0.42({\rm g-r}),
\end{equation}
where ${\rm g-r}$ is the g--r color index, given by
\begin{equation}
    {\rm g-r} = 1.02({\rm B-V}) - 0.22,
\end{equation}
where ${\rm B-V}$ is the color index.
Errors for that color correction, however, can be significant for
a very red star like Proxima, as can differences in the responses
of filters, optics, and detectors among the several 
telescopes used here.
Fortunately, ASAS observations overlap with ASAS-SN for about 3 years,
ASAS-SN $V$-band with $g$-band for about 3 months, and the several
ASAS-SN telescopes with each other, allowing very accurate cross-calibration.

For internal ASAS-SN cross-calibration, we follow the procedure described
by \citet{Irving2023}, taking the telescope with the
most measurements as our reference 
%(bi--g band--had the most--see doTelCount.out in Proxima/ZacGPRoptical), 
and then sequentially 
(in order of their measurements' temporal overlap with the reference set)
applying small offsets to other telescopes
so that the average difference among
overlapping measurements is zero. 
As each telescope is added,
the combined data set is then used as the reference. 
The same process
was applied to merge ASAS data with the ASAS-SN set, 
yielding a combined optical light curve spanning 23 years.

% In Zac's paper we fit the
% ASAS and ASAS-SN data separately, deriving Pcyc=6.7+/-1.4 years
% for ASAS and a very weakly constrained 18.3+/-14.2 yr for ASAS-SN.
% (For the latest Swift proposal you got ~11 yrs from combined X-ray and UV.)
% It would be nice to try a combined fit, but there are the complications
% of increasing brightness after ~2012 (from contamination by other stars
% as Proxima's proper motion moves it into a more crowded field) and
% a probable increase in cycle period, from ~7 years to ~11 years (judging
% from the fits to near-final X-ray/UV data used in our most recent Swift
% proposal).

%%%%%%%%%%%%%%%%%%%%%%%%%%%%%%%%%%%%%
%% FIGURE 1--Merged optical data showing weakening cyclical behavior
%%%%%%%%%%%%%%%%%%%%%%%%%%%%%%%%%%%%%
\begin{figure*}[ht!]
%\plotone{./proxima_all_data_GP_fit.pdf}
%\plotone{./OptAllData-eps-converted-to.pdf}
  \begin{minipage}[t]{0.99\linewidth}
    \begin{minipage}[t]{0.99\linewidth}
      \centering
      \includegraphics[width=0.46\linewidth]{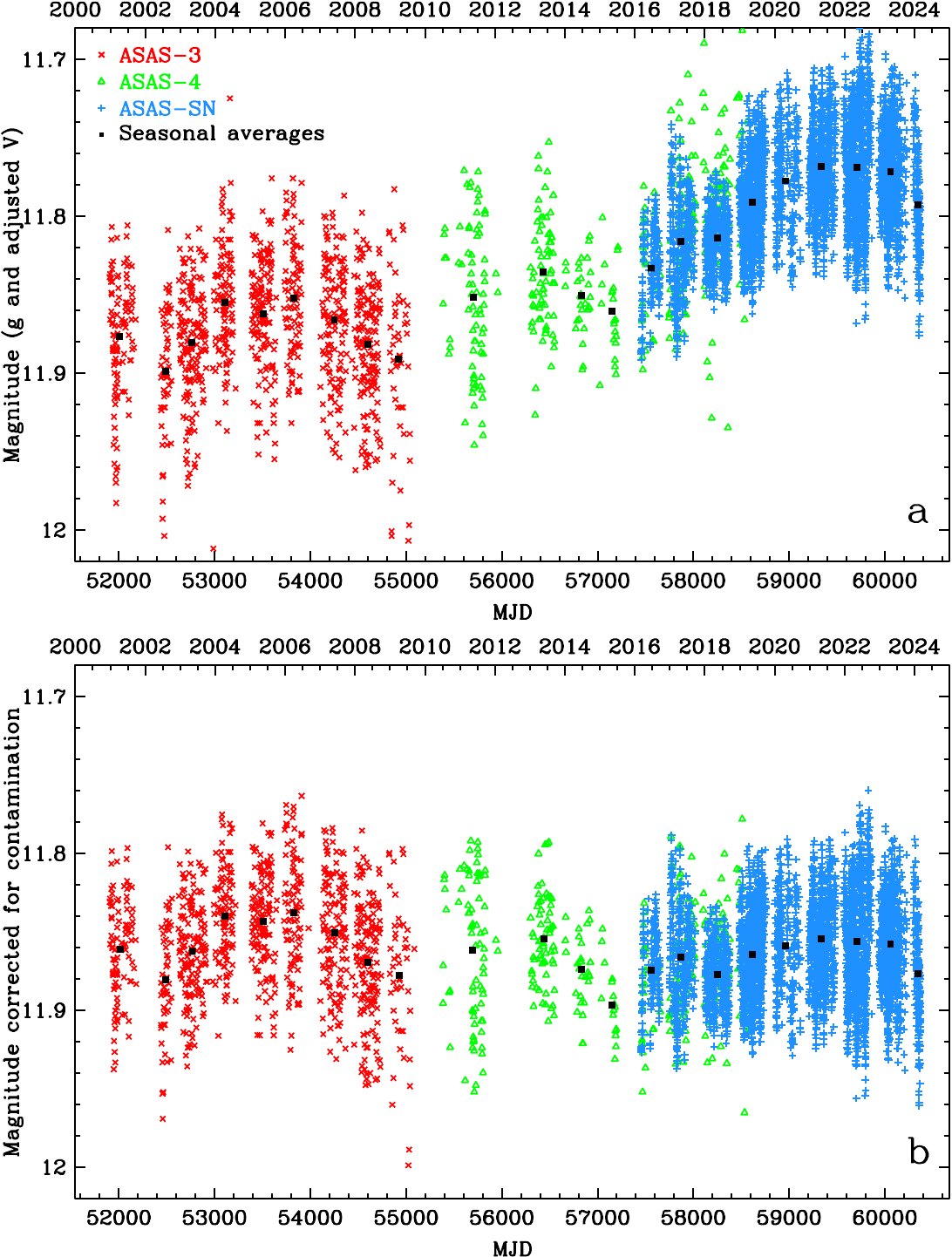}
      \hspace{3mm}
      \includegraphics[width=0.50\linewidth]{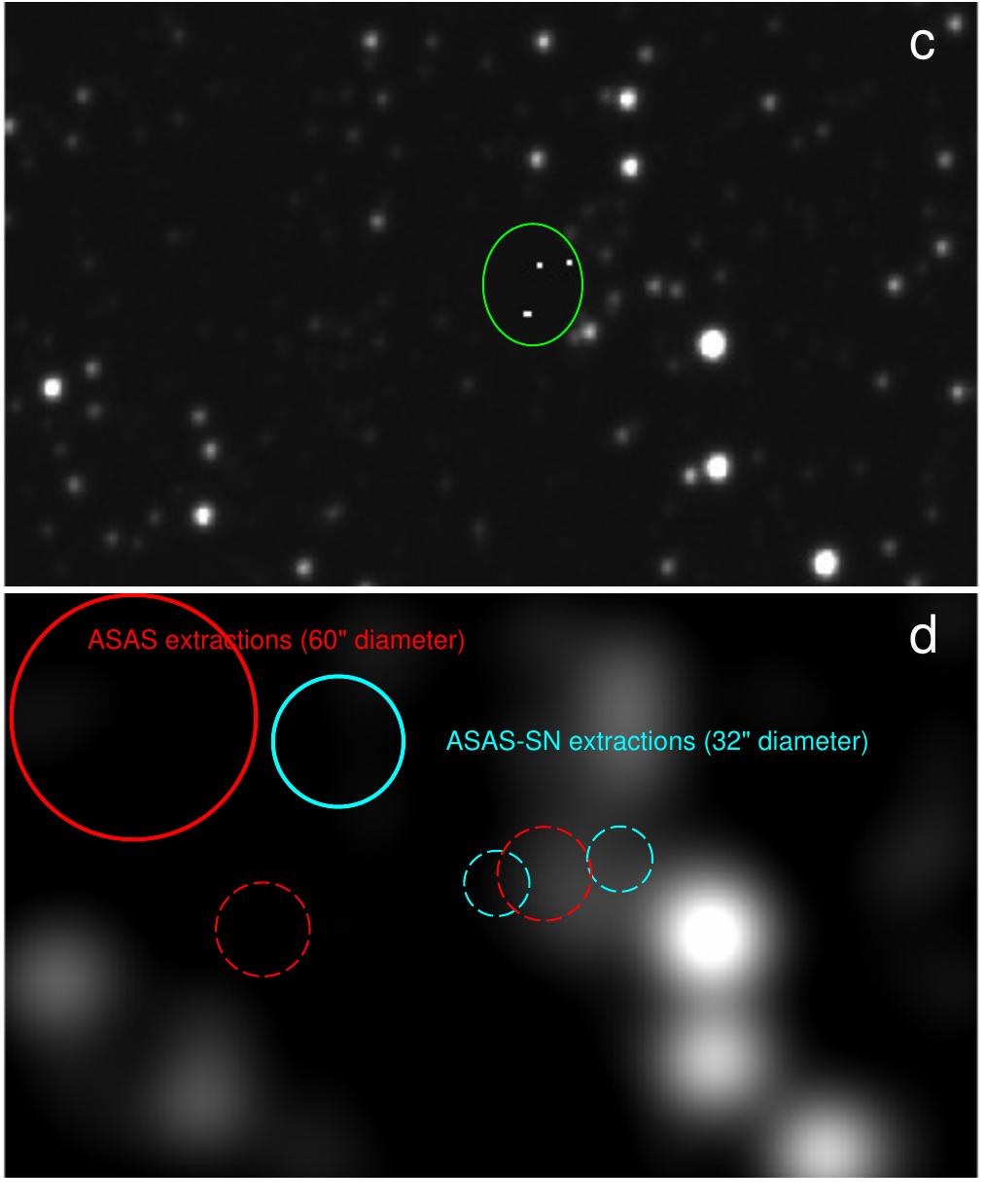}
%%% I edited the bounding box of ds9noBar.eps to start at y=-12 to make
%%% it align better with OptContamAndUN.eps.
      \vspace{0mm}
    \end{minipage}
    \centering
    \includegraphics[width=0.49\linewidth]{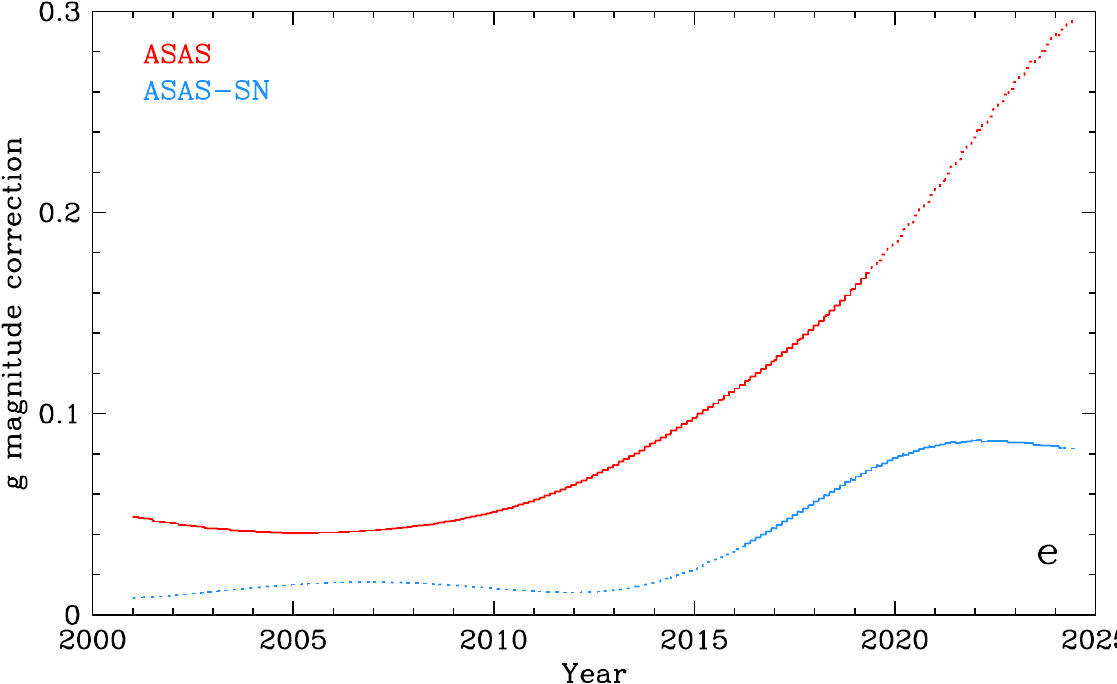}
  \end{minipage}
\caption{
Optical light curves and corrections for stellar contamination.
(a): ASAS-3, ASAS-4, and ASAS-SN data, after
cross calibration procedures described in the text.
Seasonal average for 2011 includes the few measurements from 2010 and 2012.
(b) Light curve after the correction process illustrated in panels c--e.
(c): SkyMapper $g$-band image (linear scaling) 
with Proxima removed (green ellipse) and
three obscured sources added back using data from an observation
when Proxima was at a different location.
(d): Image from (c) convolved with the ASAS 23\arcsec-FWHM PSF.
Pairs of dashed circles show the FWHM for ASAS (red) and ASAS-SN (blue) 
at Proxima's
position for the first (left) and last (right) observations in each
data set.  Solid circles in the upper left show sizes of ASAS and
ASAS-SN brightness
extraction regions.
(e): Model results for stellar contamination of Proxima, with
solid lines denoting observation intervals for both data sets.
\label{fig:OptAllData}}
\end{figure*}
%%%%%%%%%%%%%%%%%%%%%%%%%%%%%%%%%%%%%
%% An alternative would be something from Zac's paper:
%\begin{figure}
%\includegraphics[width=0.98\linewidth]{./ZacOptCycGPR-eps-converted-to.pdf}
%\caption{Alternative showing cycle fit (from Zac's paper).  
%Add the cycle curve to  Fig 1?  But we don't discuss GPR fitting
%until Section 2.4 (Rotational Modulation), add maybe Section 3
%(Stellar Cycle fitting). At the moment, I'd say add the curve and
%say we talk about it in Section TBD.  Also, we'd like the colors
%to agree with those in Fig 6, assuming we show that, which I'm
%learning toward.}
%\end{figure}
%%%%%%%%%%%%%%%%%%%%%%%%%%%%%%%%%%%%%

As seen in the first panel of Figure~\ref{fig:OptAllData}, 
optical data show the clear sinusoidal pattern
of the $\sim$7 yr cycle reported by 
\citet{SM2016} and \citet{Wargelin2017}, but in recent years
the cycle seems to have become somewhat weaker and 
there is an overall brightening trend, 
which \citet{Irving2023} suggested is caused by
Proxima's proper motion into a more densely populated region of the sky.
%which may also be obscuring the cycle to some degree. Given the optical
%telescopes' resolution of 16\arcsec\ or more, attempting
%to correct for source blending is beyond the scope of this paper.

%%%%%%%%%%%%%%%%%%%%%%%%%%%%%%%%%%%%%%%%%%%%%%%%%%%%%%%%%%%%%%%
%\subsubsection{Corrections for Stellar Contamination} 
%\label{subsec:OpticalContam}
%%%%%%%%%%%%%%%%%%%%%%%%%%%%%%%%%%%%%%%%%%%%%%%%%%%%%%%%%%%%%%%

To correct for the presumed stellar contamination we convolved an optical
image of the sky with the ASAS and ASAS-SN PSFs, simulated extractions
along Proxima's path, and adjusted the measured magnitudes.  To construct
the reference image we used $g$-band CCD observations from SkyMapper
\citep[DR4 doi:10.25914/5M47-S621;][]{Onken2024}
collected on 2015-03-09 and -10 (nominally 5 s exposures, but effectively
a little less than 4 s) and 2019-04-11 (100 s).
The longer exposure provides good S/N even for weak sources, but Proxima
itself was saturated, so its brightness versus other stars was
calibrated from the two short exposures, yielding a scaling factor of 26.5.
We removed an ellipse around Proxima in the 100 s image, replacing it with
a flat background and manually editing some pixels around the ellipse edges
to restore the PSF wings of a few adjacent stars.
In the earlier short exposures, collected 
when Proxima was 15\arcsec\ away due to 4 years of proper motion, we measured
the intensities of three stars that were obscured by Proxima in the long
exposure and restored them in the reference image, scaled up 26.5$\times$.
%(We distributed each star's counts over $3\times3$ pixels, 
%or $1.5\arcsec\times1.5\arcsec\; this doesn't match the true PSF
The resulting reference image is shown in panel c of 
Figure~\ref{fig:OptAllData}, and after subtracting a flat background it
was convolved with Gaussians of FWHM=23\arcsec\ (for ASAS; panel d)
and FWHM=16\arcsec\ (for ASAS-SN, not shown).

Counts within circles matching the ASAS and ASAS-SN extraction regions,
with diameters of 60\arcsec\ and 32\arcsec\ respectively, 
were then measured as a function
of date along Proxima's proper motion path and compared to 
the counts extracted from Proxima itself in the 5 s exposures
%Proxima's corresponding extracted counts 
%(from the average of the 5 s exposures, 
(scaled up 26.5$\times$ to match the 100 s exposure).  
The date-dependent fractional contamination of Proxima's
measured intensity caused by nearby stars was then subtracted from
the uncorrected data, and 
cross calibration of ASAS-4 and ASAS-SN in their overlapping interval
(2016--2019) was repeated, yielding
the corrected light curve in panel b of Figure~\ref{fig:OptAllData}.
Cyclic behavior is now more apparent, but still relatively muted
in recent years.  
Further discussion of the optical cycle is presented in 
Section~\ref{sec:Results},
preceded by analysis of rotational modulation in Sections~\ref{sec:GPRstuff}
and \ref{sec:OptSwiftCorrelations}.
%{\it Preceding sentence may need revision depending where all the
%GPR and L-S stuff ends up.}

%%%%%%%%%%%%%%%%%%%%%%%%%%%%%%%%%%%%%%%%%%%%%%%%%%%%%%%%%%%%%%%
\subsection{X-ray Data} \label{subsec:DataXray}
%%%%%%%%%%%%%%%%%%%%%%%%%%%%%%%%%%%%%%%%%%%%%%%%%%%%%%%%%%%%%%%

\citet{Wargelin2017} analyzed X-ray and UV data from 
the Swift Observatory
spanning 2009 to 2013, primarily
from AOs 5 and 8 with a few observations from AOs 6 and 7.  Additional
observations presented here cover AOs 12 to 17 (excluding AO 14), 2016 to 2021.
X-ray data are from the X-Ray Telescope 
\citep[XRT;][]{Burrows2005} 
and UV from the UltraViolet/Optical Telescope 
\citep[UVOT;][]{Roming2005} 
in the $\sim$1000-\AA-wide W1 band centered around
the chromospheric \mgii\ h\&k lines at 2803.5 and 2796.3 \AA.
The h\&k transitions, $3P_{1/2,3/2} \rightarrow ^3S_{1/2}$,
are analogs of the 
$4P_{1/2,3/2} \rightarrow 4S_{1/2}$
\caii\ H\&K transitions widely used
in studies of stellar cycles and other magnetic activity.
%but stronger in cool M stars like Proxima.--Actually, they might
%be stronger in most stars(?).

\citet{Wargelin2017} also included X-ray data from other missions 
back to 1994, but observations prior to 2009 were of questionable 
value because of calibration uncertainties and/or source flaring.  
The most useful measurements were made by XMM in 2009 and by the
Chandra High Resolution Camera (HRC) in 2012 and 2015.  Those and
newer measurements by the same instruments will be discussed in
Sections~\ref{subsubsec:DataXMM} and \ref{subsubsec:DataHRC}.

%%%%%%%%%%%%%%%%%%%%%%%%%%%%%%%%%%%%%%%%%%%%%%%%%%%%%%%%%%%%%%%
\subsubsection{Swift} \label{subsubsec:DataXRT}
%%%%%%%%%%%%%%%%%%%%%%%%%%%%%%%%%%%%%%%%%%%%%%%%%%%%%%%%%%%%%%%

Swift observation information is listed in Table~\ref{tab:SwiftObss}.
XRT and UVOT observations are essentially simultaneous, but their
exposures for AO 5 differ because the first eight observations
used the UVOT UV grism before switching to the W1 filter.
(The hope was to measure \mgii\ line intensities but the star field
was too crowded, often leading to overlapping dispersed spectra.)
Also, in AO 15 there was a separate Proxima observing program focused on 
flaring \citep{MacGregor2021} that provided enough
additional exposure time that AO 15 could be split into two epochs for
our analysis, but most of those observations (all of the 95159 series) used
the UVOT M2 filter instead of W1.
Observation cadences were generally multiples of 4 days 
(4, 8, 12, or 16 depending on the year)
to match the Swift ``filter of the day'' schedule, with enough observations
during each epoch (except for AOs 6 and 7)
to monitor Proxima fairly evenly over one or two stellar rotations.
%(discussed in more detail in Section~\ref{subsec:RotationalMod}).

% UVOT ultimate exposure times, including after SSS exclusions,
% are in 14SSS/sumGoodExpos.out (and sumGoodExpos.out.67 done manually)
% AO5     20938.316000000039     
% AO6     8023.51
% AO7     2684.54
% AO8     39386.811999999998     
% A12     34469.176999999901     
% A13     27637.062999999969
% A15     29382.129000000219     
% A15D    18875.029000000002     
% A16     37381.776000000114     
% A17     28530.880999999958     

\begin{deluxetable*}{ccccDD}
\tablecaption{X-ray and UV observations
\label{tab:SwiftObss}
}
\tablewidth{0pt}
\tablehead{
\colhead{}	& \colhead{}      & \colhead{}    & \colhead{}      & \multicolumn{4}{c}{Exposure time (s)\tablenotemark{b}} \\
%\vspace{-6.5mm} \\
\colhead{Mission} & \colhead{Epoch} & \colhead{ObsIDs\tablenotemark{a}} 
				& \colhead{Dates} & \multicolumn{2}{c}{X-ray} & \multicolumn{2}{c}{UVOT/W1} \\
%\vspace{-5mm} 
}
\decimals
\startdata
XMM   & 5X	& 0551120301,201,401	& 2009 Mar 10--2009 Mar 14	& 75772.0	& \nodata \\
Swift & 5	& 90215002--90215022	& 2009 Apr 23--2010 Apr 09	& 38631.0	& 20938.3\tablenotemark{c} \\
Swift & 6	& 31676001--31676003	& 2010 Jul 10--2011 Mar 12	& 7908.0	& 8023.5	\\
Swift & 7	& 31676004--31676005	& 2011 Sep 04--2011 Sep 08	& 2685.3	& 2684.5	\\
Chandra & 8C	& 14276			& 2012 Jun 15			& 49626.4	& \nodata \\
Swift & 8	& \vtop{\hbox{\strut 31676006--31676022,}\hbox{\strut 91488001--91488003}}
				& 2012 Mar 30--2013 Feb 18	& 38973.5	& 39386.8	\\
Chandra & 12Ca	& 17377			& 2015 Dec 09			& 35900.0 & \nodata \\
Swift & 12	& 31676023--31676038	& 2016 Jul 11--2016 Dec 08	& 36687.8	& 34469.2	\\
Chandra & 12Cb	& 19788-19790,19793	& 2016 Sep 26--2016 Dec 08	& 36814.2 & \nodata \\
Swift & 13	& 93156002--93156012	& 2017 Aug 25--2017 Nov 05	& 30829.8	& 27637.1	\\
XMM   & 13X	& 0801880201,401,501	& 2017 Jul 27--2018 Mar 11	& 58819.8 & \nodata \\
Swift & 15	& \vtop{\hbox{\strut 95121001--95121007,}\hbox{\strut 31676040--31676043,}\hbox{\strut 95159001--95159015}}
				& 2019 Apr 10--2019 Jul 01	& 82409.9	& 29382.1\tablenotemark{d}	\\
Swift & 15D	& 95121008--95121015	& 2019 Dec 26--2020 Feb 27	& 18877.5	& 18875.0	\\
Swift & 16	& 95673001--95673012	& 2020 Apr 08--2020 Oct 01	& 37333.7	& 37381.8	\\
Swift & 17	& 96044001--96044018	& 2021 Apr 18--2021 Aug 18	& 31020.2	& 28530.9	\\
%0551120301	2009-03-10	28	sm/med	sm/med	lg/med
%0551120201	2009-03-12	30	sm/med	sm/med	lg/med
%0551120401	2009-03-14	29	sm/med	sm/med	lg/med
%0801880201	2017-07-27@12:21:09 	2017-07-27@18:27:49 	0.00	20 ks
%0801880301 	2017-08-16@22:07:40 	2017-08-17@06:41:00	0.25	29 ks
%0801880401      2017-09-07@06:57:45	2017-09-07@14:27:45	0.50	25 ks
%0801880501	2018-03-11@20:00:59 	2018-03-12@02:40:40	0.75	23 ks
%14276 	HRC-I/NONE 	45.76 	Cal 	2012-06-15 
%17377 	HRC-I/NONE 	43.88 	Cal 	2015-12-09 
%19788 	HRC-I/NONE 	11.59 	Warg 	2016-09-26 
%19789 	HRC-I/NONE 	9.96 	Warg 	2016-10-14 
%19790 	HRC-I/NONE 	9.66 	Warg 	2016-11-03 
%19793 	HRC-I/NONE 	9.83 	Warg 	2016-12-08 
\enddata
\tablecomments{
%XRT exposures DO NOT include 3132.3 s from snapshots shorter than 200s that were
%ultimately discarded from our analysis.
%I don't think the UVOT times include short snapshots.
\vspace{-1.0mm}
\tablenotetext{a}{Swift ObsID ranges may include gaps from rescheduled
or unsuccessful observations.}
\vspace{-1.0mm}
\tablenotetext{b}{Swift times exclude periods of bad aspect,
	too-large SSS effects, or too-short exposures.
	HRC times exclude periods of telemetry saturation.
	XMM times are livetimes for the pn detector operating with 
	large window (deadtime fraction $\sim$7.5\%);
	Swift and HRC times are ONTIMEs (deadtime $<$0.5\%).}
\vspace{-1.0mm}
\tablenotetext{c}{Swift ObsIDs 90215002--90215010 used the UV grism with UVOT.}
\vspace{-1.0mm}
\tablenotetext{d}{Swift ObsIDs 95159001--95159015 used the M2 filter with UVOT.}
}
\end{deluxetable*}

%Explain about snapshots later.

All Swift data (XRT and UVOT) were downloaded from the Leicester data archive
at {\url https://www.swift.ac.uk/swift\_portal/} where they were
processed using HEASoft v6.28, particularly the FTOOLS package
\citep[{\url https://heasarc.gsfc.nasa.gov/ftools};][]{Blackburn1995}.
%Can't find a reference anywhere for HEASoft--seems more  like a wrapper
% XANADU, FTOOLS, etc.  I'm going to only give the ref for FTOOLS,
% as listed at the bottom of
% https://heasarc.gsfc.nasa.gov/docs/software/lheasoft/
%%%% Leicester's archive often uses a more
%%%% recent software version than the GSFC Science Data Center archive, and its
%%%% Data Products Generator, which we require in our analysis,
%%%% uses their local version.
% and CALDB version x20180710 (according to my notes, but the
% obs1718 header info has lines like 
% /aps/calibration/caldb_b20171016_xrt20200724_u20201215).
% obs0513 header has lines like
% Hea_15Aug2017_V6.22... and
% /aps/calibration/caldb20130405_clock112_xrt20150721
We used the Chandra Interactive Analysis of Observations (CIAO) software
suite \citep{Fruscione2006} for most analysis tasks, particularly the 
{\tt dmcopy}, {\tt dmlist}, and {\tt dmstat} commands.
For each observation we computed approximate source positions 
(with proper motion corrections) and then performed iterative centroiding 
using circular regions with 20 pixel
(47.2\arcsec) radii and 300--2500 eV energy filtering ({\tt pi=30:249}),
followed by unfiltered source and background (60--110 pixel annular) 
extractions.  The XRT PSF has a half-power radius of 9\arcsec\ and so
the source region includes some power from nearby blended sources,
but inspection of a Chandra image (ObsID 49899) with much better
resolution ($<$0.5\arcsec\ FWHM)
indicates that no more than 1\% of the extracted XRT counts
come from these other sources, even when restricted to times
of Proxima's quiescent emission.

The Swift observations last $\sim$800 s on average, with the longest
being 1850 s. Each consists of one or more ``snapshots'' 
% (up to nine in our data)
usually with intervals of one or more 95 min orbits.  We separated the
128 XRT observations into their 397 component snapshots,
further subdividing 10 snapshots into pre-flare and flare components, 
and then used the light curve function of the 
``Swift-XRT data products generator''
\citep{Evans2009}
at {\url https://www.swift.ac.uk/user\_objects/} 
to determine net 300--2500 eV source event rates 
for each snapshot (or pre/flare subdivision), 
corrected for dead pixels, 
event pileup, and other smaller effects.
%Those rates were then compared with the
%%same-energy-range 
%background-subtracted rates derived 
%from our independent source and background extractions to determine correction
%factors for each snapshot.
%(Note that a 2021 recalibration of the Swift XRT PSF increased
%corrected rates by $\sim$10\% for PC mode; SWIFT-XRT-CALDB-10\_v01.pdf.)
Those rates were then used to determine correction factors for
our own rate analysis of each snapshot, so that appropriate rates could
be calculated for an arbitrarily binned light curve, which is necessary
for our analysis (see below) but not a
capability that the products generator can directly provide.
%(Note that a 2021 recalibration of the Swift XRT PSF increased
%corrected rates by $\sim$10\% for PC mode; SWIFT-XRT-CALDB-10\_v01.pdf.)

Typical uncorrected quiescent event rates are $\sim$0.07 ct s$^{-1}$
so to obtain a statistically adequate number of counts per bin in
light curves we divided snapshots into $\sim$400-s pieces.
This also fits well with the typical $\sim$800~s snapshot exposure
and is significantly less than the typical flaring time scale
of several hundred seconds.  
To make optimal use of as much exposure time as possible we
divided each snapshot into equal pieces ranging from 
$400/\sqrt2$ to $400\!\times\!\!\sqrt2$ (283 to 566) seconds.  
%$400/\sqrt2$ (283)
%to $400\times\sqrt2$ (566) s.  
2.6\% of the total exposure time occurs in
snapshots shorter than 283 s and 1.0\% in those shorter than 200 s.
We discarded bins shorter than 200 s (3132.3 s in total), 
leaving 325356 s of XRT
exposure divided among 836 bins, averaging 389~s per bin.
%% Numbers from 3BinningXRT/calcbins.expos.out

%%%%%%%%%%%%%%%%%%%%%%%%%%%%%%%%%%%%%
%% FIGURE --Concatenated light curves, binned and rate-corrected
%%%%%%%%%%%%%%%%%%%%%%%%%%%%%%%%%%%%%
\begin{figure*}
%\plotone{./lcByCycCatted-eps-converted-to.pdf}
\includegraphics[width=0.98\linewidth]{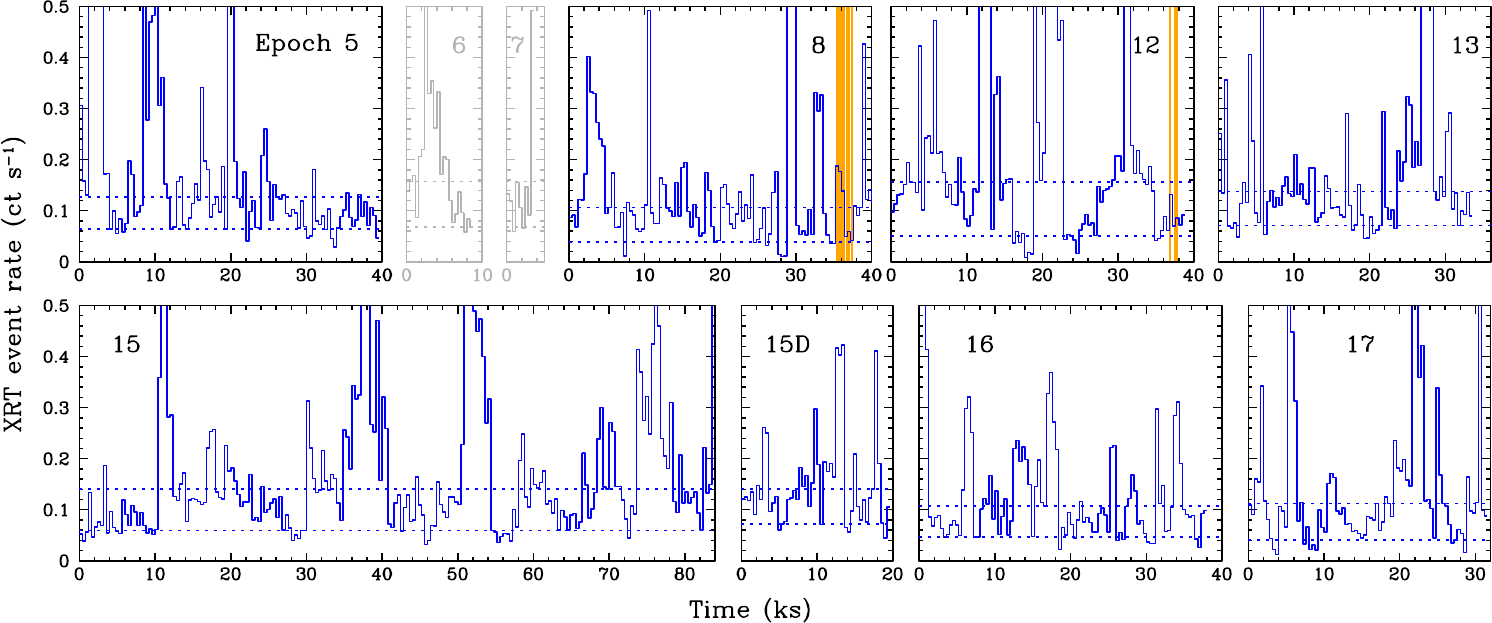}
\caption{Concatenated Swift X-ray light curves.  Time bins average $\sim$400 s
but vary so time axes are not exact. Dotted horizontal lines mark each 
epoch's 10th and 60th percentile rates 
(see Figure~\ref{fig:OrderedRateCurves}). 
Vertical orange bars denote periods of overlap with Chandra HRC observations
(see Figure~\ref{fig:XmmHrcLightCurves}).
Epochs 6 and 7 were not included in further analysis because of their 
short exposures and high proportion of flaring.
\label{fig:XrayLightCurves}}
\end{figure*}
%%%%%%%%%%%%%%%%%%%%%%%%%%%%%%%%%%%%%

% Details about XPG corrections and uncerts.
% XPG only computes statistical errors.I generously estimate the error
% c on CorrFact as 0.40*(CorrFact-1.21) with a lower limit on fractional
% c error of 1%, e.g. 1.61+/-0.16 (10% relative) and 2.01+/-0.72 (36%) and 
% c RSS that with the statistical error (sqrt#cts).
% c The fractional error on corrfact is rarely larger than statistical error--
% c a few flares (pileup and small statistical error) and sources right on or 
% c next to badpix columns. Based on plots, I do not trust corrfact values 
% c larger than ~1.9 for low rates; for flares it doesn't really matter.  
% cccccc  (Reconsider) Set raterr to 9.9999 for such cases to reflect this.
% Eh, most of the 9.999 cases are for large pileup (preflare cases get
% the same XPG corrs as the Flares--shouldn't be but I don't take the
% time now to investigate) and the few remaining seem to be very low-count
% cases where the XPG doesn't like the poor stats --> all OK.  Yes, there
% are a few non-flare in-quiescent-range cases where the correction factor 
% is >2 but not enough to worry about, and the errors should average out.

\begin{figure}
%\plotone{./newRDF1panel-eps-converted-to.pdf
%\includegraphics[width=0.98\linewidth]{./newRDF1panel-eps-converted-to.pdf}
\includegraphics[width=0.98\linewidth]{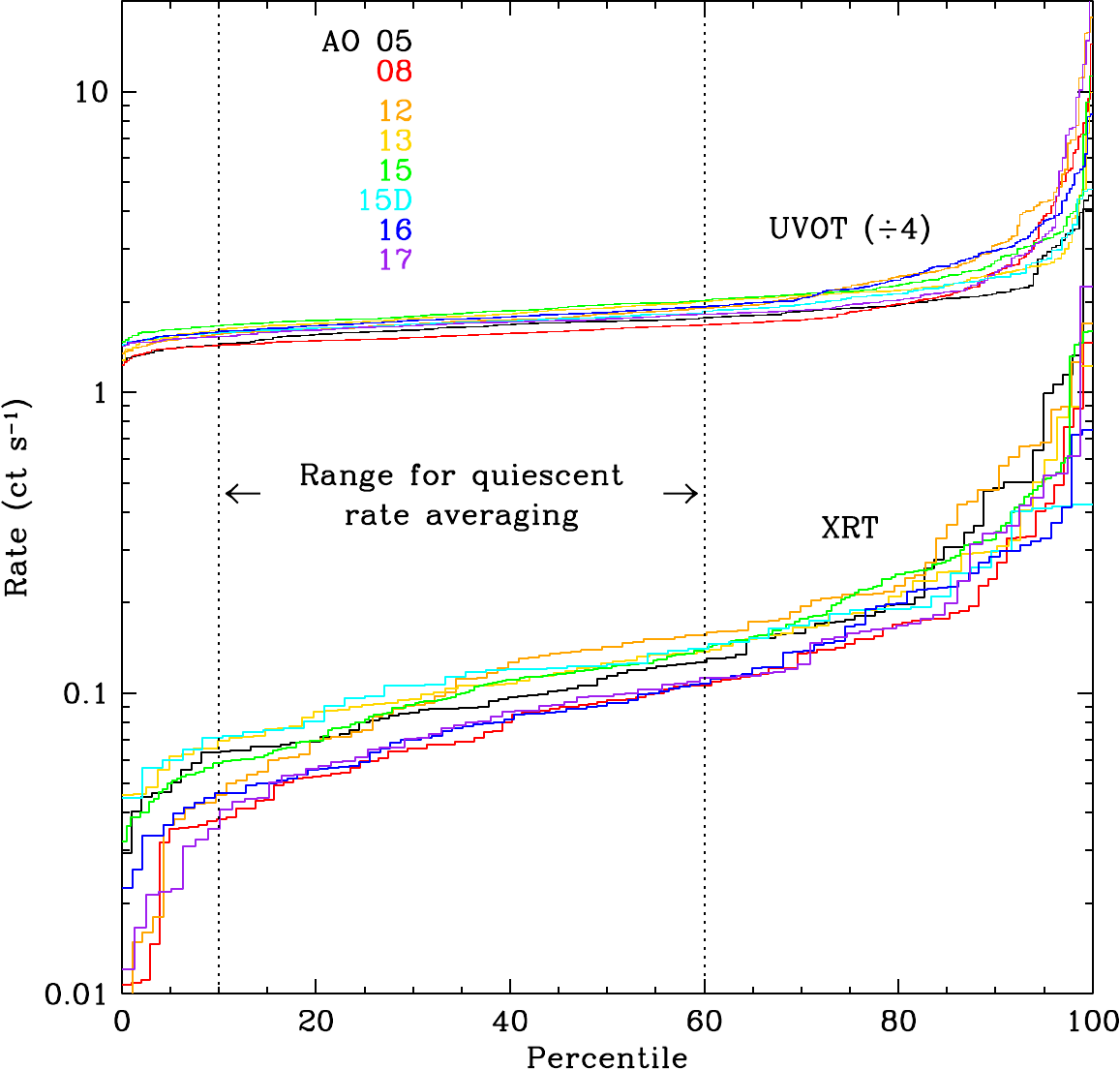}
\caption{
XRT and UVOT/W1 rate distributions for each epoch, with vertical lines marking
the 10--60th percentile range used for calculating average quiescent rates.
UVOT rates are scaled down by a factor of four in the plot.
%, and are not adjusted for rotational modulation--too small to be seen,
%  would just cause confusion to mention here.
%(see Section~\ref{subsec:RotationalMod}).
\label{fig:OrderedRateCurves}}
\end{figure}

Corrected background-subtracted rates were then computed for each bin;
Figure~\ref{fig:XrayLightCurves} shows
concatenated light curves for each epoch.
Rates during large flares are less accurate because finer time binning 
would be required to compute their highly rate dependent correction factors;
our focus is on the quiescent rates that characterize
the ``baseline'' emission of Proxima and its changes over the stellar cycle.
%We did not concentrate on obtaining accurate rate corrections for flares
%because our focus is on the quiescent rates that characterize
%the `baseline' emission of Proxima and its changes over the stellar cycle.
Of the few stars with X-ray cycle monitoring, Proxima has by far the most
flaring, and including its full light curve
%with a statistically variable 
%number of large flares, 
in such an analysis would create relatively large and
poorly estimated uncertainties
because of the variable number of large flares.  
Our measurements of quiescent rates
therefore include only time bins in the 10th to 60th percentile
of event rates, where rate distributions
from one epoch to another are most similar apart from overall normalization
(see Figure~\ref{fig:OrderedRateCurves}).
With this choice, relative rates from one epoch to another are
virtually the same (typically $\la$1\% differences)
over a fairly large percentile sampling range.
%Average rates using ranges of 10--50\% and 10--70\%
%differed from the 10--60\% average (relatively)
%by typically only $\sim$$\pm$1\%, and 3\% in the worst case.
%10-70 ave rates were 9.5% larger than 10-60.
%10-50 ave rates were 7.2% smaller than 10-60.

%%%%%%%%%%%%%%%%%%%%%%%%%%%%%%%%%%%%%
%% FIGURE TBD--Merged optical data showing weakening cyclical behavior
%%%%%%%%%%%%%%%%%%%%%%%%%%%%%%%%%%%%%
%%%%%%%%%%%%%%%%%%%%%%%%%%%%%%%%%%%%%

% Somewhere around here or below explain that AOs6+7 didn't have enough 
%exposure time, seems to mostly be coming off flares, not well spaced, etc.

Source and background event files from each of those time bins were merged
into composite quiescent emission files for each epoch, and their spectra fit
with Sherpa \citep{Freeman2001}
using the detector Ancillary Response File
({\tt swxs6\_20010101v001.arf}) 
and appropriate Response Matrix File (RMF) for each epoch 
({\tt swxpc0to12s6\_20090101v014.rmf} for AO 5, 
{\tt *20110101v014.rmf} for AO 8, 
and {\tt *20130101v014.rmf} for AO 12 and later).
There was too little exposure time in AOs 6 or 7, even when combined,
to obtain a reliable estimate of their quiescent emission level,
especially since their snapshots' light curves indicate 
most of their limited exposure time probably occurred during significant
flaring (see Figure~\ref{fig:XrayLightCurves}).

We used the Sherpa
{\tt xsphabs} model in combination with either two or three
{\tt xsapec} components at different temperatures for our fits.  
The {\tt xsphabs} H column
density was frozen at $10^{18}$ cm$^2$ and had no effect in the
0.3--5 keV band used for fitting.  Two temperatures generally
gave formally acceptable results in terms of the reduced statistic,
but 3-T fits were visibly better.
Three temperatures also follows the rationale of 
%\citet{Favata2008},
\citet{Orlando2017} and 
\citet{Coffaro2020} in their analyses
of X-ray stellar cycles in HD 81809
and $\epsilon$ Eri, respectively, that three $kT$'s are necessary to model
the emission from active regions (AR; lowest $kT$),
AR cores (CO; medium $kT$), 
and higher temperature regions that produce flaring
emission that is subtle enough not to be obvious in light curves
(FL; highest $kT$).
A fourth component at lower temperatures 
representing plasma like that in the ``quiet Sun'' contributes negligibly to 
detected X-ray emission from these stars and is not included.

%%%%%%%%%%%%%%%%%%%%%%%%%%%%%%%%%%%%%
%% FIGURE--Sherpa fits (3T) to Cycles 8 and 12
%%%%%%%%%%%%%%%%%%%%%%%%%%%%%%%%%%%%%
\begin{figure}
  \begin{minipage}[t]{1.00\linewidth}
      \centering
      \includegraphics[width=0.96\linewidth]{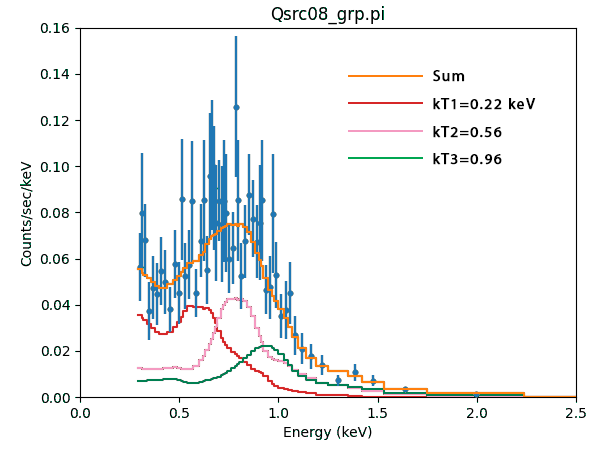}
      \includegraphics[width=0.96\linewidth]{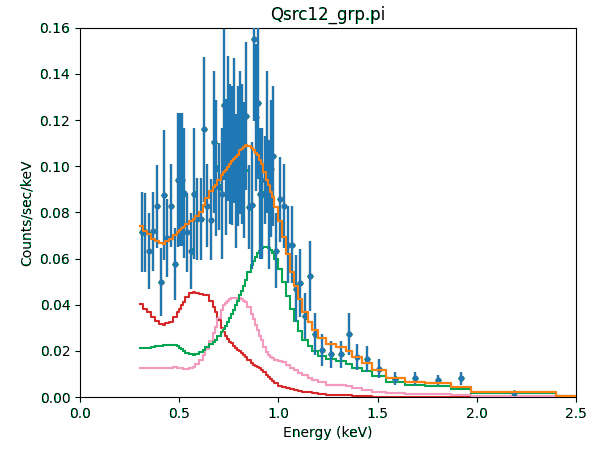}
  \end{minipage}
\caption{
Sherpa fits to quiescent X-ray data from epochs 8 and 12 (near cycle
minimum and maximum, respectively) with three
components at fixed temperatures corresponding to 
active regions,
active region cores,
and subdued flaring regions.
Note the large difference in 
high-temperature emission between the two epochs.
\label{fig:SherpaFits}}
\end{figure}
%%%%%%%%%%%%%%%%%%%%%%%%%%%%%%%%%%%%%

Following the 3-T paradigm, we fit each epoch's spectrum with frozen
temperatures and abundances, allowing only the normalization of
each temperature component to vary 
(see examples in Figure~\ref{fig:SherpaFits}).  
The fixed temperatures and abundance were taken from fits to
the combined spectra of all epochs that could be fit with a single
%RMF, namely for AO 12 and later: 
RMF (that for AOs 12-17): 
%$kT_1=0.22$ keV ($T=2.55{\rm e}6$ K),
%$kT_2=0.56$ keV ($T=6.50{\rm e}6$ K),
%$kT_3=0.96$ keV ($T=11.14{\rm e}6$ K),
$kT_1=0.22$ keV ($T=2.55\times10^{6}$ K),
$kT_2=0.56$ keV ($T=6.50\times10^{6}$ K),
$kT_3=0.96$ keV ($T=11.14\times10^{6}$ K),
and an abundance of 0.17 times solar.
% 1 kelvin [K] = 8.61732814974056E-05 electron-volt [eV] 
When leaving the abundance thawed in individual fits for each epoch, 
values ranged from 0.10 to 0.24 times solar photospheric values
with typical $1\sigma$ uncertainties of 
$\pm0.05$.  There were not enough counts to meaningfully
investigate how abundances might vary among elements with different
first ionization potentials, but the overall low
abundance for Proxima's corona is consistent with previous
findings for quiescent emission in, e.g.,  
\citet{Lalitha2020}
($0.23^{+0.86}_{-0.12}$ from Chandra LETG observations in 2017,
and $0.38^{+0.31}_{-0.16}$ from contemporaneous AstroSat measurements),
and in
\citet{Wargelin2017}
(0.25 from XMM-Newton observations in 2009).
%and \citet{Fuhrmeister2011}
%(a range of 0.30 to 0.79 for various abundant elements

\begin{deluxetable*}{cccccccc}[t]
\tablecaption{Fit results for quiescent X-ray emission
\label{tab:XrayFits}
}
\tablewidth{0pt}
\tablehead{
\colhead{Epoch} 
    & \colhead{Total}	
             & \colhead{$kT_1$} 
                      & \colhead{$kT_2$} 
                               & \colhead{$kT_3$}
                                        & \colhead{Source}
                                               & \colhead{Average}
                                                         & \colhead{Quies.} \\
%\vspace{-6.5mm} \\
\colhead{} 
    & \colhead{Flux}	
             & \colhead{fraction} 
                      & \colhead{fraction} 
                               & \colhead{fraction}
                                        & \colhead{Counts}
                                               & \colhead{MJD}
                                                         & \colhead{Exp.~(s)} \\
%\vspace{-5mm}
}
%\decimals
\startdata
%AO totalflux  f1frac   f2frac   f3frac SrcCts    AveMJD   QuietExp
% From 7FitXquiet/0FinalFluxes.corr---corrected for badpix, etc.
5X  & $2.336$         & 0.480 & 0.258 & 0.263 & 48621 & 54902 & 37962 \\
05  & $1.952\pm0.096$ & 0.414 & 0.305 & 0.282 & 1369 & 55167 & 19213 \\
8C  & $1.22\pm0.24$   & \multicolumn{3}{c}{Fixed to Epoch 08 values} & 6602 & 56093 & 24317 \\
08  & $1.564\pm0.075$ & 0.420 & 0.356 & 0.224 & 1089 & 56083 & 19840 \\
12Ca & $4.55\pm1.07$  & \multicolumn{3}{c}{Fixed to Epoch 12 values} & 4809 & 57365 & 17551 \\
12  & $2.235\pm0.113$ & 0.322 & 0.239 & 0.438 & 1497 & 57630 & 18048 \\
12Cb & $2.24\pm0.51$  & \multicolumn{3}{c}{Fixed to Epoch 12 values} & 3332 & 57682 & 17889 \\
13  & $2.133\pm0.118$ & 0.337 & 0.286 & 0.377 & 1229 & 58029 & 15134 \\
13X & $2.460$         & 0.482 & 0.291 & 0.227 & 40539 & 58053 & 29563 \\
15  & $2.104\pm0.071$ & 0.416 & 0.237 & 0.347 & 2967 & 58620 & 41158 \\
15D & $2.168\pm0.151$ & 0.345 & 0.322 & 0.332 &  768 & 58861 &  9482 \\
16  & $1.614\pm0.081$ & 0.455 & 0.178 & 0.367 & 1049 & 59038 & 18502 \\
17  & $1.618\pm0.086$ & 0.376 & 0.292 & 0.332 &  957 & 59381 & 16266 \\
\enddata
\tablecomments{
Fluxes are for $E=0.3-2.5$ keV in units of $10^{-12}$ erg cm$^{-2}$ s$^{-1}$
(equivalent to a luminosity of $2.0283\times10^{26}$ erg s$^{-1}$
at Proxima's distance).
%Uncertainties are scaled in proportion to the inverse square root
%of the quiescent exposure time such that 20 ks yields 5\% error.
Fluxes are adjusted for vignetting and enclosed energy fractions and 
(for Swift) missing pixels
but do not include adjustments for rotational modulation
(see Section~\ref{subsec:RotationalMod} and 
Table~\ref{tab:UVOTaveRates}).  
Uncertainties for Chandra HRC fluxes (the ``C'' epochs) represent
normalization statistics only (see text). 
XMM flux uncertainties (``X'' epochs) from
counting statistics are very small; systematic uncertainties are difficult
to estimate but larger than for Swift measurements.
% Nah, this is only for the weird srcflux results: Fractional contributions 
% of each $kT$ component were held fixed to Swift-derived values for 
% HRC flux determinations.
}
\end{deluxetable*}
% AO  flux1 flux2 flux3 total error f1frac f2frac f3frac SrcCts AveMJD QuietExp
% 05  8.075 5.948 5.500 19.523  0.957 0.4136 0.3047 0.2817 1369 55167.1 19213.2
% 08  6.573 5.572 3.498 15.643  0.754 0.4202 0.3562 0.2236 1089 56083.1 19839.8
% 12  7.209 5.350 9.793 22.351  1.130 0.3225 0.2394 0.4381 1497 57630.3 18047.5
% 13  7.190 6.102 8.037 21.330  1.178 0.3371 0.2861 0.3768 1229 58029.2 15134.4
% 15  8.752 4.991 7.301 21.045  0.705 0.4159 0.2372 0.3469 2967 58619.8 41158.4
% 15D 7.485 6.993 7.203 21.681  1.512 0.3452 0.3225 0.3322  768 58861.2  9482.2
% 16  7.343 2.865 5.932 16.141  0.806 0.4549 0.1775 0.3675 1049 59038.0 18502.4
% 17  6.089 4.726 5.363 16.178  0.862 0.3764 0.2921 0.3315  957 59381.1 16265.7
% Fluxes are corrected for badpix, etc.
% Uncerts scale as 1/sqrt(exposure time) so that 20ks
% will give 5 percent errors.
% See 0FinalFluxes for uncorrected data + other info

Using the above temperatures and abundance
in fits to individual epochs, we obtained
the results in Table~\ref{tab:XrayFits}, which have been adjusted for 
enclosed energy fractions and missing pixels
by applying exposure-weighted averages of 
the previously calculated XRT rate corrections,
ranging from 1.291 (AO 12) to 1.378 (AO 15).  
%There is no way to adust
%the event data for missing pixels prior to fitting, so we applied correction
%factors to each epoch's fitted flux, derived from the exposure-weighted
%average of the previously calculated corrections for each 
%time bin, which ranged from 1.291 (AO 12) to 1.378 (AO 15).
Statistical uncertainties from the number of counts in each spectrum
are $\sim$3\%, which we believe are relatively unimportant compared
to uncertainties arising from sampling of Proxima's highly
variable emission. 
Estimating the ``true'' uncertainties of the measured fluxes is
a fraught task, but longer total exposure times
will generally provide a more accurate measure than shorter times.
We therefore assigned uncertainties proportional to $1/\sqrt{T_{{\rm exp}}}$, 
with 20 ks of quiescent exposure scaled to yield 5\% error.
The main importance of those errors is providing the relative weights
used in fitting the stellar cycle, as discussed in
Section~\ref{sec:Results}.

Converting the fitted fluxes in Table~\ref{tab:XrayFits} to luminosities,
we see that quiescent emission (0.3--2.5 keV) ranges from 
3.2 to $4.5\times10^{26}$ erg s$^{-1}$, agreeing well (keeping
in mind that our criteria for quiescence are generally stricter)
with levels found by \citet{Lalitha2020} ($3.5\times10^{26}$
from LETG, and $5.1\times10^{26}$ from AstroSat),
\citet{Fuhrmeister2011} ($5.0\times10^{26}$ from XMM-Newton in 2009),
and \citet{Gudel2004} (4--$28\times10^{26}$ from XMM in 2001).
The table also shows that higher total emission is associated
with a higher fraction of emission from the hottest ($kT_3$) component,
which is particularly well illustrated in Figure~\ref{fig:SherpaFits}.

%%%%%%%%%%%%%%%%%%%%%%%%%%%%%%%%%%%%%%%%%%%%%%%%%%%%%%%%%%%%%%%
\subsubsection{XMM} \label{subsubsec:DataXMM}
%%%%%%%%%%%%%%%%%%%%%%%%%%%%%%%%%%%%%%%%%%%%%%%%%%%%%%%%%%%%%%%

XMM first observed Proxima in 2001 but flaring made those data
unsuitable for measuring the quiescent emission level.
A set of three observations in March 2009 was analyzed in 
\citet{Wargelin2017}, and their quiescent emission level agreed well with
that from the close-in-time Swift AO5 epoch.  We have reanalyzed 
those data from the EPIC pn detector 
along with a set of four newer observations from 2017 and 2018 
(see Table~\ref{tab:SwiftObss}).  ObsID 0801880301 in the latter
set consisted entirely of strong flare emission and was excluded
from further analysis.

We downloaded data from the XMM archive and reprocessed using
the SAS software (version xmmsas\_20230412\_1735-21.0.0) with standard
pn filtering.  The source was extracted using a circle with 400 pixel
(20\arcsec) radius, with background ($\sim$0.5\% of the source rate)
taken from a region of the same size offset by 100\arcsec.  Using 200 s bins,
we applied the same quiescent emission criteria as for the Swift analysis,
created rate filters using the {\tt tabgtigen} tool to generate
the quiescent event files, and then made RMF and ARF files using
{\tt rmfgen} and {\tt arfgen}.

The resulting spectra were fit using Sherpa with the same 3-temperature
{\tt xsphabs} model used in the Swift XRT analysis.  Abundance was
likewise fixed at 0.17; freeing it resulted in values of 0.165 and 0.148
for epochs 5X and 13X, respectively, with negligible flux changes.
The fitted fluxes, listed in Table~\ref{tab:XrayFits}, are a little
higher than contemporaneous Swift results.  
%In the limit of continuous
%observing, Swift and XMM should yield the same results, but finite
%sampling 
XMM has longer exposure times for each epoch than Swift, but
because rates in each time bin are not truly independent, particularly
during flares, Swift's snapshot sampling will tend to produce a more
representative and generally slightly lower rate distribution than XMM 
for a given total exposure time.
Net uncertainties on the XMM fluxes 
%are probably about the same as for Swift, but 
are not listed in Table~\ref{tab:XrayFits} because it is not clear
how their values relative to Swift should be computed.

%%%%%%%%%%%%%%%%%%%%%%%%%%%%%%%%%%%%%
%% FIGURE --Concatenated light curves, binned and rate-corrected
%%%%%%%%%%%%%%%%%%%%%%%%%%%%%%%%%%%%%
\begin{figure*}
\includegraphics[width=0.98\linewidth]{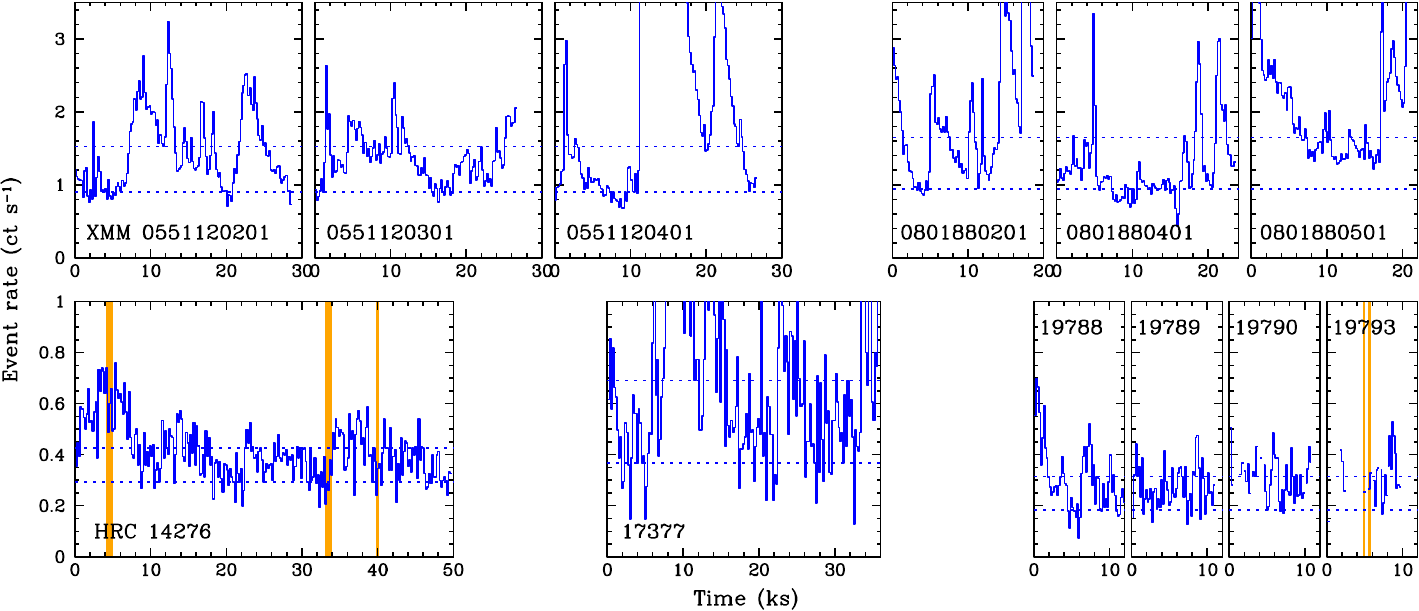}
\caption{X-ray light curves for XMM pn (epochs 5X and 13X) and 
Chandra HRC (epochs 8C, 12Ca, and 12 Cb) observations.
Dotted horizontal lines mark quiescent rate limits.
Vertical orange bars denote periods of overlap with Swift observations
(see Figure~\ref{fig:XrayLightCurves}), allowing flux normalizations.
The first overlap was not used for normalization
because it was during a flaring period.
\label{fig:XmmHrcLightCurves}}
\end{figure*}
%%%%%%%%%%%%%%%%%%%%%%%%%%%%%%%%%%%%%

%%%%%%%%%%%%%%%%%%%%%%%%%%%%%%%%%%%%%%%%%%%%%%%%%%%%%%%%%%%%%%%
\subsubsection{Chandra HRC-I} \label{subsubsec:DataHRC}
%%%%%%%%%%%%%%%%%%%%%%%%%%%%%%%%%%%%%%%%%%%%%%%%%%%%%%%%%%%%%%%
%The first two observations
%were single auxiliary calibration measurements made far off-axis
%(15.0\arcmin\ and 25.62\arcmin, respectively), and the set
%of four 2016 observations were also made at 15.0\arcmin\ to minimize
%systematic uncertainties.

Chandra's High Resolution Camera for Imaging (HRC-I) microchannel plate
detector observed Proxima in 2012, 2015, and 2016.  All measurements
were made off axis in ``Next In Line'' mode with restricted telemetry; 
our analysis removed periods of telemetry saturation caused by
background flares, which occurred at
the beginning and end of the 2015 observation, and during several 
intervals in the last
two of the four observations in 2016 (see Figure~\ref{fig:XmmHrcLightCurves}).
There were also short periods in 2012 and 2016 when Swift and the HRC
observed Proxima simultaneously, allowing rates observed by the HRC,
which has essentially no energy resolution, to be normalized against
fluxes detected by Swift.

%% DTF values are junk in NIL mode.  After removing the periods of
%% telem saturation, deadtime is darn close to 0 so I can just use ONTIME.

For the 2012 and 2016 observations, made 15.0\arcmin\ off axis,
the source was extracted using an ellipse with radii of 207 and 315 pixels
(27.3\arcsec\ and 40.2\arcsec),
background from a surrounding elliptical annulus of equal area.  The 2015
observation, made 25.62\arcmin\ off axis, used source radii of 
460 and 740 pixels and a surrounding background annulus.
Periods of quiescent emission were again identified using 10th to 60th 
percentile rates, although without energy filtering, and quiescent event files
were constructed for each of the three epochs.

Spectral fitting cannot be used for HRC data so we used the CIAO
{\tt srcflux} tool.  We specified the same source models used
for Swift and XMM fitting, except that the relative normalizations
of the three temperature components were necessarily fixed, leaving
only the overall flux as a free parameter.  For the HRC 8C epoch we used
the $kT$ fractions from Swift epoch 8 (see Table~\ref{tab:XrayFits}),
and Swift epoch 12 fractions for 12Ca and 12Cb.  Swapping the 
distributions changed the derived fluxes by only $\sim$8\%.

{\tt srcflux} results are for the HRC energy range of 0.1--10 keV,
so to estimate the 0.3-2.5 keV flux we used the Sherpa 
{\tt calc\_energy\_flux} tool on Swift fitting results from
epochs 8 and 12, deriving scaling factors of 0.675 and 0.692, respectively.
The resulting fluxes for epochs 8C, 12Ca, and 12Ca are then
6.10, 11.53, and $4.74\times10^{-12}$ erg s$^{-1}$ cm$^{-2}$,
which are all much larger than the fluxes measured by Swift or XMM.
It is difficult to understand these results, but they are presumably
related to the very different energy dependent responses of the
HRC microchannel plate detector versus the Swift and XMM CCDs,
a subject discussed by \citet{Ayres2008} to explain the initially
% Ayres_2008_ApJ_678_L121.pdf , "The Fainting of α Centauri A, Resolved"
puzzling ``fainting'' of $\alpha$ Cen A.
The CCD detectors of Swift and XMM have effectively no useful response
below $\sim$300 eV because of electronic noise, but the HRC
response extends to about 100 eV and therefore detects emission
within the C-K transmission window below 284 eV
(see Figure~12 of \citet{Wargelin2017}).
  
The short periods of simultaneous observation 
by Swift and the HRC, however, provide
the opportunity for direct cross calibration; 
see the orange time 
bands in Figures~\ref{fig:XrayLightCurves} and \ref{fig:XmmHrcLightCurves}.
The first of three periods of overlap 
during Swift epoch 8 and HRC epoch 8C/ObsID 14276 in 2012 occurred
when the source was flaring and we therefore do not use it in our
normalizations for quiescent emission;
the total overlap during the other two intervals was 1103 s.
The total during the 2016 overlap 
(Swift epoch 12 and HRC epoch 12C/ObsID 19793) was 458 s.
% 14276/0822bc overlap is 441.29+441.29+220.64=1103.22 s.
% 19793/1238b overlap is 110.0+347.95=457.95 s.

Since the quiescent Swift rates and fluxes have already been measured,
we can use HRC vs Swift rates during the times of overlap
to convert the HRC quiescent rates to fluxes with
\begin{equation}
F_{Hq} = R_{Hq}\frac{R_{Ho}}{R_{So}} \frac{F_{Sq}}{R_{Sq}}
\end{equation}
where $F$ is flux and $R$ is measured event rate, and the subscripts
$H$, $S$, $o$, and $q$ denote HRC, Swift, overlap, and quiescence,
respectively.
Applying the 2012 normalization to HRC epoch 8C and
the 2016 normalization to epochs 12Ca and 12Cb,
we obtain fluxes of
($1.22\pm0.24$),
($4.55\pm1.07$),
and $(2.24\pm0.51)\times10^{-12}$ erg s$^{-1}$ cm$^{-2}$
where the listed uncertainties are from only the normalization statistics
and do not include systematic errors.

The middle measurement (epoch 12Ca) is clearly an outlier,
and the light curve in Figure~\ref{fig:XmmHrcLightCurves} suggests
that observation was dominated by flaring with only a few very brief
periods of quiescent emission.
The measurement from epoch 12Cb should be the most reliable, since
its four separate observations come closest to approximating the
nearly random sampling of Swift's many snapshots.  Epoch 8 (ObsID 14276)
comprises only one observation, but it is relatively long, does
not have the strong flares seen in ObsID 17377, and includes substantial
periods with rates near the overall minimum, so we believe it is
probably a good indicator of the quiescent emission level.

Excluding the highly suspect epoch 12Ca, 
the cross-calibrated HRC results are indeed consistent with
those from Swift (see Table~\ref{tab:XrayFits}.
We note, however, that the normalization factors ${R_{Ho}}/{R_{So}}$
for epochs 8 and 12 differ by a factor of $2.697\pm0.807$,
a surprisingly large difference.
Like the puzzling {\tt srcflux} results, this is probably due
to significant differences in spectral distributions during the
two epochs, presumably at low energies that the Swift CCD is not sensitive to.

There are two other sets of X-ray observations since 
the work done in \citet{Wargelin2017}, four with Chandra HRC-S/LETG
in 2017, and two with ACIS-S/HETG in 2019.  The use of the LETG and HETG
gratings, however, decreases rates by nearly an order of magnitude, and 
ACIS/HETG has very little of its sensitivity below $\sim$1 keV.  
Because of their very low count rates, we do not include these
observations in our work here.
%Given the
%very low count rates and other limitations of these instruments'
%data when studying Proxima's stellar cycle,
%we do not include them in this work.

%%%%%%%%%%%%%%%%%%%%%%%%%%%%%%%%%%%%%%%%%%%%%%%%%%%%%%%%%%%%%%%
\subsection{UVOT Data} \label{subsec:DataUVOT}
%%%%%%%%%%%%%%%%%%%%%%%%%%%%%%%%%%%%%%%%%%%%%%%%%%%%%%%%%%%%%%%

Analysis of Swift UVOT data is similar to that for the XRT, but with additional
steps to correct for contamination by nearby stars, and for temporal and
spatial variations in quantum efficiency (QE).  
After splitting each observation into its
component snapshots, we extracted 240$\times$160 pixel (120.5\arcsec$\times$
80.3\arcsec) ellipses around Proxima to reduce file sizes by 96\% for subsequent
steps.  Source count rates are roughly 100 times
higher than in the XRT so we used $\sim$100 s binning instead of $\sim$400 s
to divide up the snapshots, which is short enough that there was no
need to manually separate preflare and flare intervals.  A few snapshots
were removed or time-filtered because
Good Time Intervals were incorrect
(usually because the spacecraft had not finished slewing),
%(removed 15D28c, 0704c, 1303e, 1701b, others?; 
the target fell outside the field of view,
% (one snapshot),
%Prox is outside UVOT FoV for 15C17a;
%time-filtered 1223d,1701a) 
or exposures were too short ($<100/\sqrt(2)$ s).

%A study of flares, however, could be interesting.  There's at least one
%case where there's a signif UVOT flare but no XRT flare, vice versa,
%and everything in between.
%
%Anomalousrates.info,pdf has XRT very low rate followed by flare for
%   0818ab (UVOT has low rate [4.75 ct/s] for a and big decaying flare for b)
%   1233bPF (can't tell offhand if UVOT rate was initially subpar)
%and to a lesser degree for
%   1235b (nothing obviously odd about b/c in UVOT)

%%%%%%%%%%%%%%%%%%%%%%%%%%%%%%%%%%%%%%%%%%%%%%%%
%%%% Possible subsubsection---Extraction Regions
%%%%%%%%%%%%%%%%%%%%%%%%%%%%%%%%%%%%%%%%%%%%%%%%

The intrinsic UVOT resolution is 2.5\arcsec\ FWHM \citep{Breeveld2010} but
there is usually a little pointing drift during observations,
%and aspect corrections are not perfect, 
with errors during snapshots sometimes
reaching 4\arcsec.  We can recapture the intrinsic
resolution by centroiding the source in each 100 s time bin, which enables
tighter extraction regions to be used resulting in less contamination
by nearby sources.  Consistent source sizes also permit more accurate
corrections for source contamination when it is unavoidable, which
is a larger concern than with X-ray observations because Proxima
does not completely dominate its local field in the UVOT/W1 band as it
does in X-rays, and its cycle amplitude in W1 is smaller.

Our choice of UVOT source extraction radius was informed by the
Swift website's ``UVOT Aperture Correction'' discussion\footnote{
{\url 
https://swift.gsfc.nasa.gov/analysis/threads/uvot\_thread\_aperture.html}}.
S/N is maximized with extraction radii equal to the PSF FWHM,
%(2.5\arcsec,
%or 3\arcsec\ to allow for temperature effects, etc.),
but Proxima is relatively bright and
we are more interested in the relative accuracy of event rates, which primarily
means reducing sensitivity to centroiding errors and
PSF variations.  We therefore chose a larger extraction radius of 5\arcsec\
(10 pixels), which allows for shifts in the source center of $\pm1$ pixel
with negligible change in the derived source rate.
For background, we used the recommended annulus with
radii of 27.5\arcsec\ and 35\arcsec\ centered on the source.
Event pile-up effects are negligible for source rates of 
$<$10 ct s$^{-1}$, 
comfortably above Proxima's typical quiescent rates of 6--8 ct s$^{-1}$.

As noted in Section~\ref{subsec:DataOptical},
Proxima's proper motion is carrying it through a crowded region of
the sky,
% and optical monitoring surveys with resolutions of $\sim$20\arcsec\
% are unable to avoid blending.  
and even with UVOT's resolution of 2.5\arcsec,
some source contamination is unavoidable.
In order to correct for this, we stacked images of all W1
observations to improve S/N and quantify the brightness of each nearby source,
and then determined the fraction of each star's emission that fell
within Proxima's (moving) source extraction region.

%Need to stack images to get better S/N on weak sources to determine their
%average rates and positions as a function of date relative to Proxima.

%Determine pixel centroids (iterative dmstat)
%of Prox and its PM-RA/DEc position for each bin.
%11.SumIMG/ImgSumFromArcObs/ contains summed images using pipeline/archival 
%images of each observation.
%I created my own summed images, stacking images from centroided 100 s slices
%by using do100centroids.out results and calculating Proxima's proper-motion-
%corrected RA,Dec for each slice and physical coord offset from a reference
%point (the center of the images to be created) in alignimages.f to create
%the dmcopy/binning commands in doextractaligned.
%I then used doSum100IMG.* scripts to sum images and create sum100Cyc*.img.
%I will use those images (or maybe just sum100Cyc12to13.img, for which
%Proxima doesn't overlap with any other sources) in 12ExtSrcAndBG/ to
%determine how to extract source and BG evt files.

\begin{figure}
\includegraphics[width=0.98\linewidth]{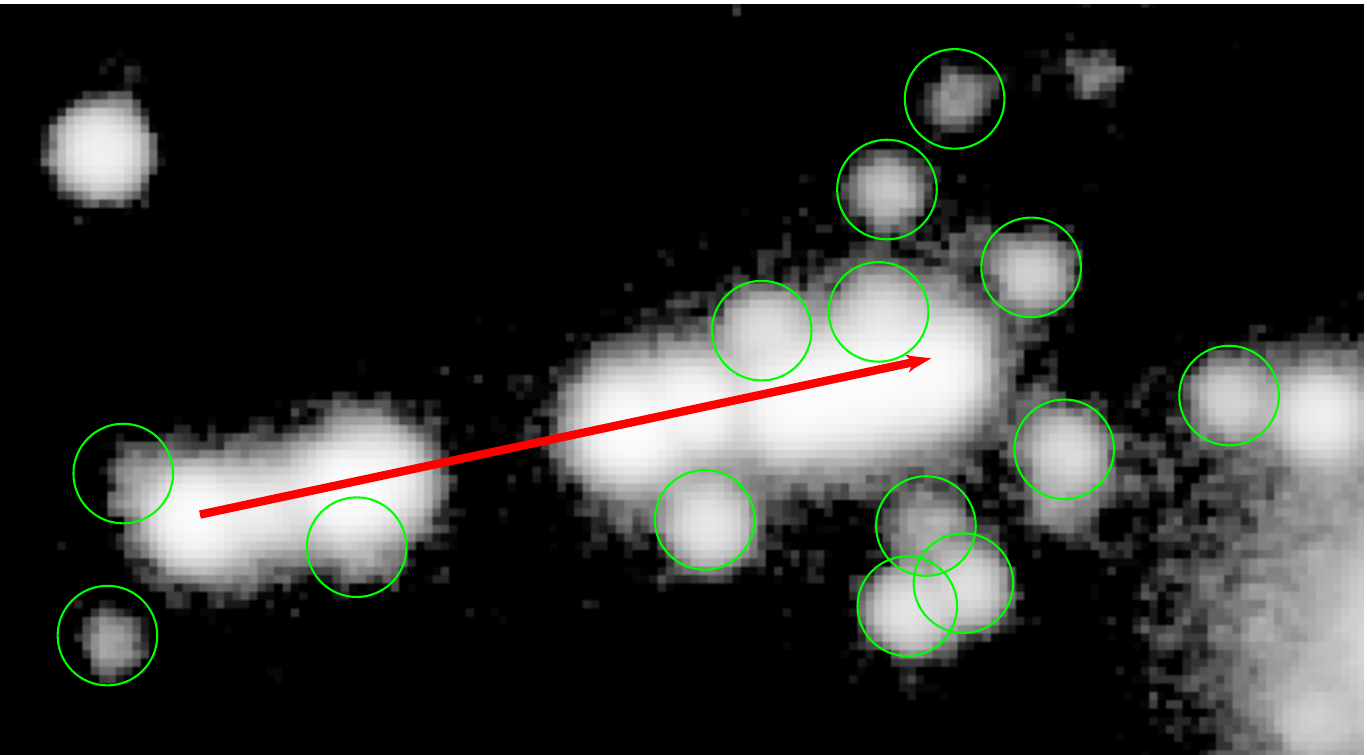}
\caption{
Summed UVOT/W1 data (highly stretched) from AOs 5 to 17. 
Proxima's proper motion (47.56\arcsec) over that time interval 
is shown in red. 
Green circles (radius of 6 pixels, or 3\arcsec) mark sources
that can interfere with Proxima or its background-region annuli;
UVOT PSF is $\sim$2.5\arcsec\ FWHM.
Rate corrections for interference were all below 2\%.
\label{fig:UVstackedIMG}}
\end{figure}

To align images for stacking, we first
computed Proxima's well determined date-dependent RA and Dec for each snapshot
and measured its iteratively centroided source position (in pixel
coordinates) in each 100 s time bin.
An image file was then created for each time bin, centered on pixel coordinates
corresponding to a common reference position in RA and Dec, near the center
of Proxima's motion over 12 years; alignment of each image was 
accurate to $\pm$0.5 pixels in both dimensions, limited by the necessity
to use integer pixel coordinates for image extractions.
%
%I want to extract an image file for each time bin that is centered as
%precisely as possible on the same reference sky position (RA and Dec).
%As noted above, the evt file RA and Dec coordinates are typically only
%good to a couple arcsec so I use the pixel-coordinate centroids of
%Proxima (iterative dmstat; accurate to of order 0.1 pixels), 
%for which we compute proper-motion-corrected sky 
%coordinates (accurate to better than 0.1\arcsec), 
%to compute the image extraction pixel coordinates that
%will yield the same field of view for each time bin (to within 0.5 pixels;
%image extractions necessarily use integer pixel coordinates).
% dmcopy 'ell0511a.fits[time=274182264.607:274182366.158,(x,y)=box(1978,2007,400,300)][bin x=::1,y=::1]' ell0511a01.img
Finally, the resulting images were summed using {\tt dmimgcalc} 
into composite files
for individual and multiple epochs.  The combined image of all
observations is shown in Figure~\ref{fig:UVstackedIMG}.  
Analysis showed that only Proxima has detectable proper
motion so the other stars' positions can be treated as fixed.
%(12ExtSrcAndBG/sum5to17wReg.jpg
%or 12ExtSrcAndBG/sum5to17wRegCROP.eps which is from the Swift prop.
%I think the former uses 10 pix radii.)

%%%%%%%%%%%%%%%%%%%%%%%%%%%%%%%%%%%%%
%% FIGURE TBD
%%%%%%%%%%%%%%%%%%%%%%%%%%%%%%%%%%%%%
%%%%%%%%%%%%%%%%%%%%%%%%%%%%%%%%%%%%%

%Using those combined images and the calculated RA and Dec for Proxima,
%the absolute positions all stars that could interfere with Proxima's
%source or background regions were calculated, and their separations
%from Proxima at any time.  Relative intensities 
%
%and relative positions of 
%Those combined images were used to determine the relative position
%and average count rate
%of all stars that could interfere with Proxima's source or
%background extraction regions. 
%Their
%positions relative to Proxima as a function of time can then be calculated,
%along with the fraction of each star's counts within Proxima's 10 pixel-radius
%source region, which was calibrated separately using a summed stellar 
%image and extraction circle with variable center-to-center distance.
%(**Explain better?)

From those combined images we determined the absolute position
and average count rate of all stars that could interfere with 
Proxima's source or background regions. 
Their positions relative to Proxima during each observation were
then calculated, along with the fraction of each star's counts 
that would fall within Proxima's 10 pixel-radius source region. 
That fraction was calibrated separately using a summed stellar 
image and measuring the extracted counts
while varying the distance between the 
%center of the
star and 
%the center of 
the extraction circle.
Corrections for source contamination were $\sim$2\% for
AOs 15, 16, and 17, $<$1\% for AOs 8 and 13, and $<$0.4\% for AOs 5 and 12.
Uncertainties from those corrections, assuming a 1 pixel error in
Proxima's source centroid, were all $<$1\%.
Proxima's annular background regions were re-extracted when necessary to exclude
10 pixel regions around any contaminating sources,
and net rates were adjusted for
the slow decline in UVOT/W1 QE (about 30\% over the last 17 years;
CALDB file {\tt swusenscorr20041120v006.fits}),
%% Yes, the QE changes are a bit different for different bands.
using QE at the beginning of 2005 as the baseline.

%%%%%%%%%%%%%%%%%%%%%%%%%%%%%%%%%%%%%%%%%%%%%%%%
%%%% Possible subsubsection---Small Scale Sensitivity Corrections
%%%%%%%%%%%%%%%%%%%%%%%%%%%%%%%%%%%%%%%%%%%%%%%%

One final UVOT QE consideration is the
Small Scale Sensitivity (SSS) issue\footnote{ 
{\url https://swift.gsfc.nasa.gov/analysis/uvot\_digest/sss\_check.html}}, 
in which a small fraction
of the detector area has significantly lower QE than the rest.
In the W1 band the QE can be up to 17\% low
over several percent of the central 5\arcmin$\times$5\arcmin\
of the field
(SWIFT-UVOT-CALDB-17-02$\,$\footnote{
{\url https://heasarc.gsfc.nasa.gov/docs/heasarc/caldb/swift/docs/uvot/uvotcaldb\_sss\_02b.pdf}}).
We wrote a program that reads the 
{\tt swulssens20041120v003.fits} CALDB file and produces
images of where SSS regions fall with respect to
our source and background extractions \citep{slavin.2024.13356340},
and found 12 snapshots with affected source regions.
% Yes, snapshots, not time bins.
We discarded only the five of them that had more than 5\% of their counts 
in the SSS areas, corresponding to no more than 1\% rate suppression.
SSS affected 22 snapshots' background regions, 
but only five of them over more than 5\%
of their area.  For those five, this would introduce rate errors
of no more than 0.1\%,
%background-subtracted rates would be no more than 0.1\% high,
but we conservatively re-extracted the background regions, excluded the SSS zones,
and recomputed the net rates.

In the end, 102 time bins (3.9\% of the total) 
were removed because of SSS within Proxima's
source extraction regions, leaving 2486 bins,
and 143 of those bins required adjustments to their background regions.
As was done for the XRT, we then
selected bins with rates between the 10th and
60th percentiles to determine the quiescent rate averages for each epoch
(see Figure~\ref{fig:OrderedRateCurves}
and Table~\ref{tab:UVOTaveRates}).
Similarly to how uncertainties were assigned to X-ray fluxes,
we set errors to scale as $1/\sqrt{T_{{\rm exp}}}$, with
20 ks of quiescent exposure yielding 2\% error.

For completeness, we note that the UVOT UV filters (W1 and W2) have
non-negligible transmission beyond their central wavelength ranges\footnote{
\url https://swift.gsfc.nasa.gov/analysis/uvot\_digest/redleak.html}.
To estimate contamination in the W1 band, we rescaled a quiescent UV grism
spectrum presented in Figure 4 of \cite{Wargelin2017} (primarily taken
from three snapshots in observation 009, with small pieces from other  
observations to fill in wavelength gaps) using the effective areas for the 
grism and W1 filter from the CALDB, and calculated that $\sim$16\% of the 
observed W1 signal comes from wavelengths with $\lambda > 3300$ \AA.  
As noted in that figure, an increasing fraction of the observed grism 
spectrum arises from higher orders toward longer wavelengths, so the 
actual contamination is less.  Its main effect is to increase the    
bias level of the observed W1 signal and therefore slightly reduce the 
apparent relative amplitude of W1 cycle variations.

\begin{deluxetable}{ccccc}[t]
\tablecaption{Rates for quiescent UVOT/W1 emission
\label{tab:UVOTaveRates}
}
\tablewidth{0pt}
\tablehead{
\colhead{Epoch} 
    & \colhead{W1}	
             & \colhead{Quiesc.} 
                      & \colhead{Rot.~Mod.} 
                               & \colhead{X-ray Rot.} \\
%\vspace{-6.5mm} \\
\colhead{} 
    & \colhead{Rate}	
             & \colhead{Exp.~(s)} 
                      & \colhead{Adj.~(\%)} 
                               & \colhead{Adj.~(\%)} \\
%\vspace{-5mm}
}
%\decimals
\startdata
%AO   W1rate+/-err    Expos   RotAdj   XrotAdj 
05  & $6.54\pm0.18$ & 10493 & 0       & 0 \\
08  & $6.19\pm0.24$ & 19745 & 0       & 0 \\
12  & $6.87\pm0.24$ & 17297 & $-0.19$ & $+0.56$ \\
13  & $7.19\pm0.23$ & 13830 & $+0.26$ & $-0.42$ \\
15  & $7.34\pm0.24$ & 14743 & $+0.44$ & $-2.04$ \\
15D & $6.82\pm0.18$ &  9453 & $+0.85$ & $+4.11$ \\
16  & $7.02\pm0.26$ & 18626 & $+0.42$ & $-2.69$ \\
17  & $6.74\pm0.22$ & 14297 & $-1.17$ & $-2.90$ \\
\enddata
\tablecomments{
W1 rates include corrections for contamination by nearby stars
and time-dependent QE;
rotational modulation adjustments are listed separately,
for both W1 and X-ray.
(Epochs 5 and 8 did not have enough contemporaneous optical
data to analyze modulation effects.)
%Rotational modulation adjustments for X-ray fluxes (not included in
%Table~\ref{tab:XrayFits}) are also listed.
W1 rate uncertainties are scaled in proportion to the inverse square root
of the quiescent exposure time such that 20 ks yields 2\% error.
}
\end{deluxetable}
\subsection{Correlations and Corrections: Rotational Modulation} 
\label{subsec:RotationalMod}
%%%%%%%%%%%%%%%%%%%%%%%%%%%%%%%%%%%%%%%%%%%%%%%%%%%%%%%%%%%%%%%

In addition to uncertainties in our determination of quiescent
rates arising from the intrinsically variable nature of emission
from a flare star, which we addressed by using many short observations
and 10-60th percentile sampling,
there is also the issue of systematic error caused by biased
sampling of the stellar surface.
Emission from the star is not spatially uniform---which 
is of course the reason for the rotational modulation seen 
in the optical light curve 
(Figure~\ref{fig:OptAllDataGPRalt})---and our 
Swift observations may be capturing a nonrepresentative
sample of that emission despite a fairly even cadence during each
epoch and typically monitoring
over at least one full rotation period.  
As described below, we therefore studied the correlation of
optical, UV, and X-ray emission variations to see if adjustments
are needed for the Swift measurements, as well as determine
%A study of correlations between
%optical, UV, and X-ray emission variations will indicate if adjustments
%need to be made to the Swift measurements, as well as tell us
if spot darkening or faculae brightening is the primary driver of
optical modulation.
%(How did *** that other paper explain this?)

%%%%%%%%%%%%%%%%%%%%%%%%%%%%%%%%%%%%%%%%%%%%%%%%%%%%%%%%%%%%%%%%
\subsubsection{Gaussian Process Smoothing of Optical Data}
\label{sec:GPRstuff}
%%%%%%%%%%%%%%%%%%%%%%%%%%%%%%%%%%%%%%%%%%%%%%%%%%%%%%%%%%%%%%%%

%Swift observations are much too sparse to directly detect
%rotational modulation, but we can ...
To create a smooth continuous optical light curve that can be
used to estimate optical brightness during any of the
comparatively few Swift observations,
we use Gaussian process regression \citep[GPR;][]{Rasmussen2006}{}{}
on the data shown in Figure \ref{fig:OptAllData},
implementing our GP models using the 
\texttt{tinygp}\footnote{\url{https://github.com/dfm/tinygp}} Python package.
GPR requires specification of a mean and covariance function, 
both of which should capture observed or expected trends within the data. 
Our combined light curve contains several thousand data points, 
and GPR scales as $N^3$, so a thorough comparison of different models
would be prohibitively time consuming. 
Instead, we attempt to make informed or otherwise justifiable choices to produce a model that interpolates between the data in an insightful way.

\begin{figure*}
    \centering
    \includegraphics[width=0.98\linewidth]{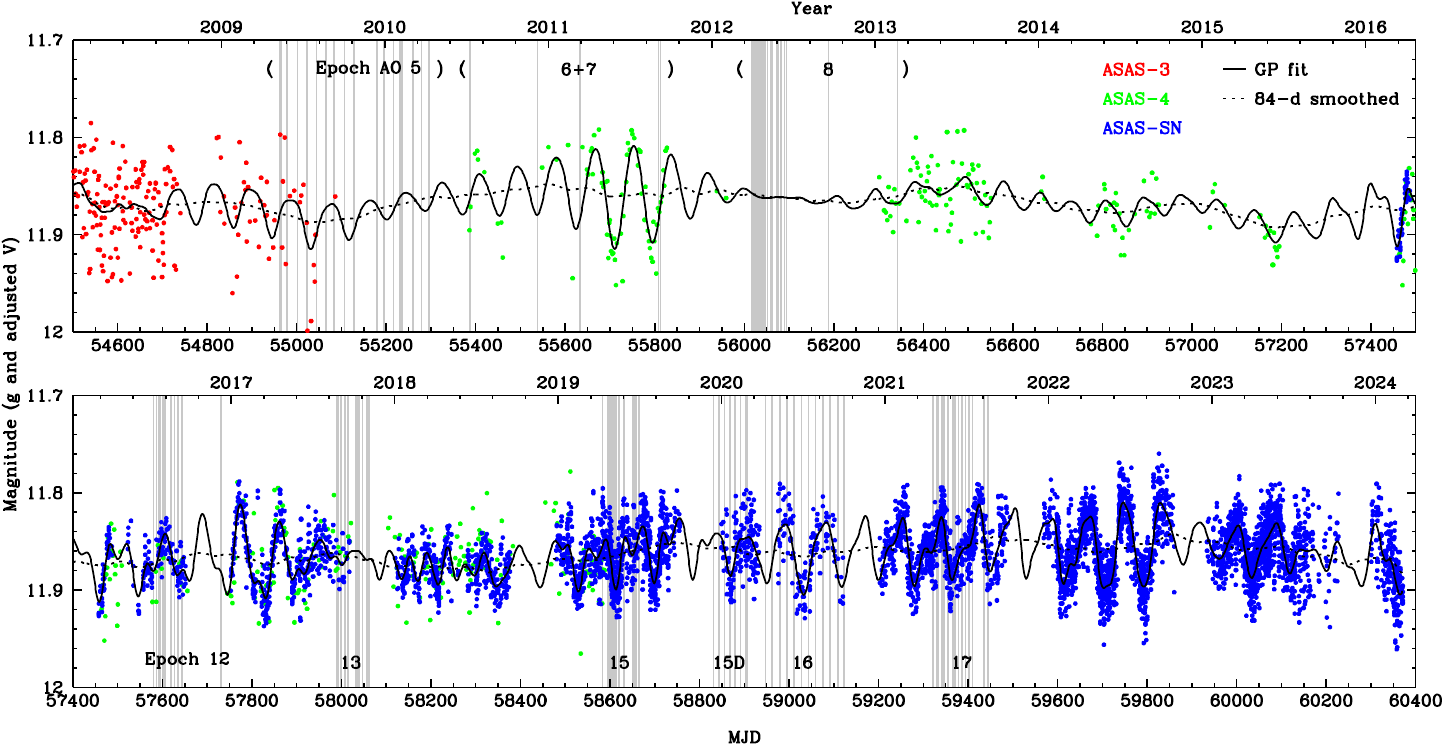}
    \caption{
%        Optical data with GPR fit showing rotational modulation,
%        and the dates of Swift observations (gray shading).
        Optical data (with corrections for stellar contamination)
        and GPR fit showing rotational modulation,
        with the dates of Swift observations marked by gray shading.
        Dotted curve is from a sliding boxcar average of the GPR fit
        over the rotation period;
        subtraction of the dotted curve yields the rotational ``residuals''
        used to study correlation with X-ray and UV variations
        (see Figure~\ref{fig:SwiftOptCorrelate}).
    \label{fig:OptAllDataGPRalt}}
\end{figure*}

For the mean function, about which the stellar emission fluctuates,
we use a simple constant:
\begin{equation}
    m(t) = c,
    \label{eq: GP mean}
\end{equation}
where $t$ is time, and $c$ is the constant. 
%We use this mean function since we expect the stellar emission to 
%fluctuate about some quiescent level, $c$. 
When choosing the covariance function, we considered the following: 
1) Proxima has a rotation period of $\sim$84 d, 
with rotational modulations clearly visible in 
some parts of the light curve
%the cross-calibrated ASAS-SN light curve 
(Figure \ref{fig:OptAllDataGPRalt}); 
2) Proxima exhibits a (quasi-)periodic stellar cycle 
\citep[][]{SM2016, Wargelin2017}{}{}; 
%3) recent observations of Proxima likely suffer significant contamination 
%from nearby stars (Figure~\ref{fig:UVstackedIMG}).
3) residual stellar contamination may still be present
after the corrections illustrated in Figure~\ref{fig:OptAllData}.
%recent observations of Proxima suffer significant contamination 
%from nearby stars, some of which may still be present after our corrections.
With these points in mind, we use a two-component covariance function. 
The first component is a slightly modified version of the quasi-periodic 
(QP) covariance function described in \cite{Nicholson2022}:
\begin{equation}
    k_{\rm QP}(t, t') = \sigma^2 {\rm exp} \Big[ - {\rm sin}^2 \Big( \frac{\pi |t\!-\!t'|}{P} \Big) - \Big( \frac{|t\!-\!t'|^2}{2 \ell^2} \Big)\Big],
% Put - in front and remove one of the two others
% You can't put the - in front because it's a power
%    k_{\rm QP}(t, t') = -\sigma^2 {\rm exp} \Big[{\rm sin}^2 \Big( \frac{\pi |t\!-\!t'|}{P} \Big) + \Big( \frac{|t\!-\!t'|^2}{2 \ell^2} \Big)\Big],
    \label{eq: quasi-periodic covariance function}
\end{equation}
where $\sigma^2$ is the variance, 
$P$ is the period of the oscillation, 
and $\ell$ is the ``length-scale.'' 
%\cite{Nicholson2022} showed that 
If $P$ is set to the 
stellar rotation period, then $\ell$ relates to the starspot evolution time, 
making this covariance function well-suited to modeling the rotational 
modulations of active stars. Equation \ref{eq: quasi-periodic covariance function} differs from the QP kernel described in \cite{Nicholson2022} by a $\Gamma$ factor, which we have implicitly assumed to be 1. The $\Gamma$ factor, sometimes called the ``harmonic complexity'' parameter, is difficult to interpret physically, but larger values produce oscillations that deviate more from a simple sinusoid. In our experience, this parameter is often highly degenerate, and oscillations can still deviate significantly from a simple sinusoid when $\Gamma = 1$. For these reasons, we omit the $\Gamma$ factor.

%%%%%%%%%%%%%%%%%%%%%%%%%%%%%%%
%\begin{figure*}
%    \centering
%    \includegraphics[width=\textwidth]{./optical_GP_fit.pdf}
%    \caption{(Pretty, but too much duplication of the 2-panel figure
%that follows? Add longer-period/cycle curve?)
%Optical data with GPR fit of rotational modulation.}
%    \label{fig:OptAllDataGPR}
%\end{figure*}
%%%%%%%%%%%%%%%%%%%%%%%%%%%%%%%

%%%%%%%%%%%%%%%%%%%%%%%%%%%%%%%
%%%%%%%%%%%%%%%%%%%%%%%%%%%%%%%

The second component is intended to
capture any variability unassociated with Proxima's rotation.
For this we use the squared 
exponential covariance function, sometimes referred to as the 
radial basis function or Gaussian kernel,
\begin{equation}
    k_{\rm SE}(t, t') = \sigma^2 {\rm exp} \Big( - \frac{|t - t'|^2}{2 \ell^2} \Big),
    \label{eq: squared exponential covariance function}
\end{equation}
where $\sigma$ and $\ell$ share the same definitions as in equation \ref{eq: quasi-periodic covariance function}. The hyper-parameters of the squared exponential are again difficult to interpret physically, but it is well-suited to modeling smoothly varying stochastic processes. 

With our GP specified, we then used nested sampling 
\citep[][]{Skilling2004, Skilling2009}{}{} to construct posterior probability 
distributions for our (hyper-)parameters, implemented with the 
MLFriends algorithm \citep[][]{Buchner2016, Buchner2019}{}{} using the 
\texttt{UltraNest}\footnote{\url{https://johannesbuchner.github.io/UltraNest/}}
\citep[][]{Buchner2021}{}{} package. 
We used \texttt{UltraNest}'s \texttt{ReactiveNestedSampler} with 
\texttt{min\_ess=1000} to yield at least 1000 posterior samples, 
and all other parameters left to their default values. 
Table \ref{tab: GPR priors} lists the prior probability distributions 
for our (hyper-)parameters. 

\begin{table}
    \centering
    \caption{(Hyper-)parameter Prior Probability Distributions}
    \begin{tabular}{ccc}
        \hline
        \hline
        (Hyper-)parameter & Equation & Prior \\
        \hline
        $c$ & \ref{eq: GP mean} & $\mathcal{U}$(min($f$), max($f$)) \\
        $\sigma$ & \ref{eq: quasi-periodic covariance function}, \ref{eq: squared exponential covariance function} & log $\mathcal{U}$(min($\sigma_f$), $10 \Delta f$) \\
        P & \ref{eq: quasi-periodic covariance function} & $\mathcal{U}$(80, 90) \\
        $\ell$ & \ref{eq: quasi-periodic covariance function}, \ref{eq: squared exponential covariance function} & log $\mathcal{U}$(min($\delta t$), $\Delta t$) \\
        \hline
    \end{tabular}
    \label{tab: GPR priors}
    \tablecomments{$\mathcal{U} (a, b)$ denotes a uniform prior covering the interval $a$ to $b$ and log $\mathcal{U} (a, b)$ denotes a prior that is logarithmically uniform. $f$ represents the flux, $\sigma_f$ represents the errors on the fluxes, and $t$ represents time. We use $\delta$ to denote the minimum (non-zero) difference between any two variables, and $\Delta$ to denote the maximum difference.}
\end{table}

Finally, we used the maximum likelihood sample to obtain the 
GP posterior in 
%Figures~\ref{fig:OptAllDataGPR}
Figure~\ref{fig:OptAllDataGPRalt}
showing Proxima's rotational modulations, which
can deviate considerably from a simple sinusoid. 
The posterior for the rotation period hyper-parameter, $P$, 
is $85.4 \pm 0.7$~d, essentially the same as the $85.1 \pm 1.2$~d period 
reported by \cite{Irving2023} but with a slightly smaller uncertainty.

%%%%%%%%%%%%%%%%%%%%%%%%%%%%%%%%%%
\subsubsection{Optical-X-ray/UV Correlations}
\label{sec:OptSwiftCorrelations}
%%%%%%%%%%%%%%%%%%%%%%%%%%%%%%%%%%

%Swift observations are not frequent enough to easily detect rotational
%modulation in the UV W1 or X-ray bands, but we can compare their emission
%levels with the much better sampled optical light curves to determine if
%and how strongly they are correlated.  To reduce noise in the optical
%data and fill in gaps in coverage (with more reliability for short
%gaps, and less for long) we perform GPR fits tuned to roughly the
%$\sim$84 d rotation period (**refs needed?--give details--see Fig ***).  
%From the GPR curve we can then estimate
%Proxima's optical magnitude during any Swift observation.

\begin{figure*}
  \begin{minipage}[b]{0.99\linewidth}
\centering
    \includegraphics[width=0.61\linewidth]{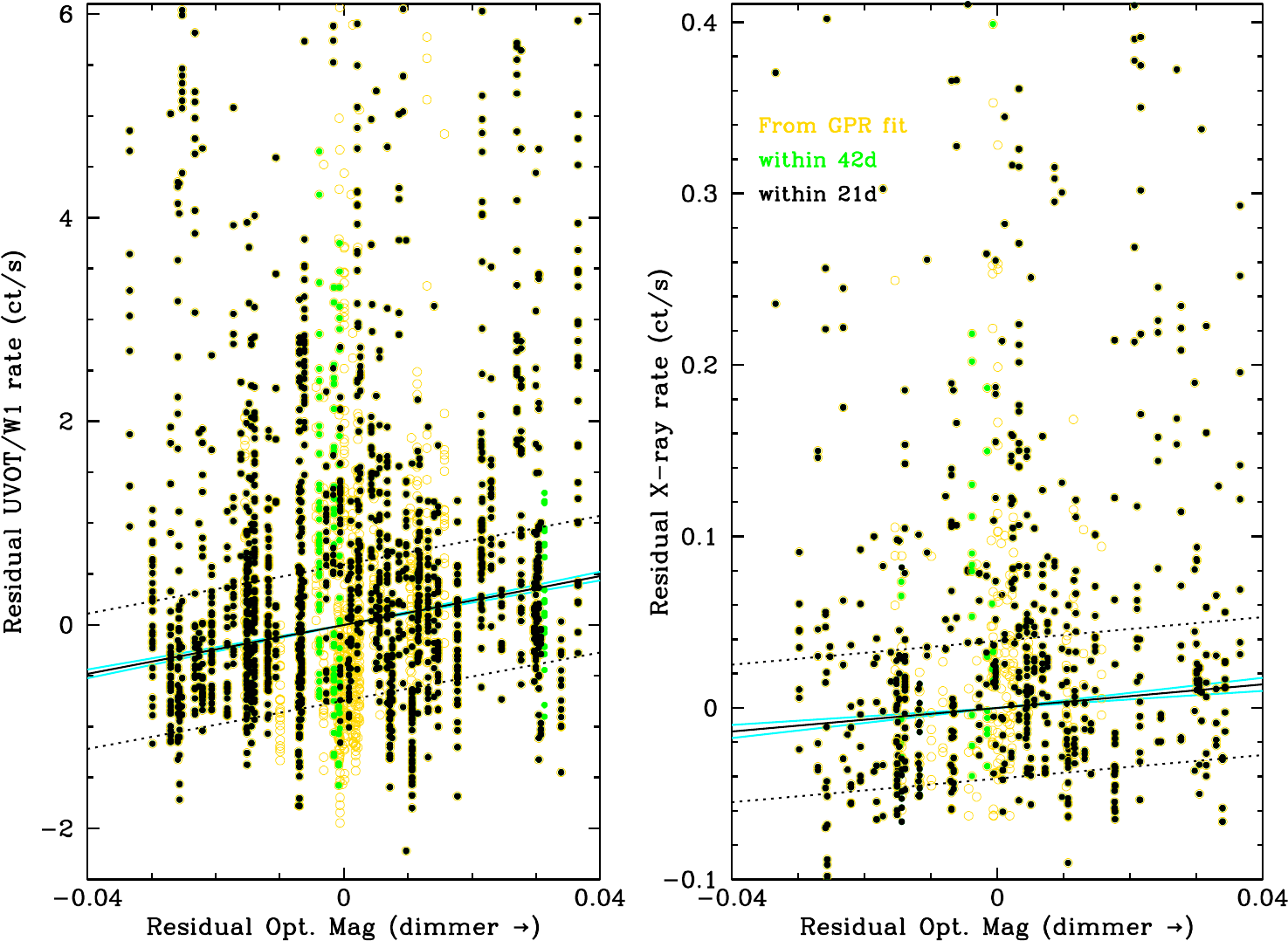}
    \hspace{5mm}
    \includegraphics[width=0.30\linewidth]{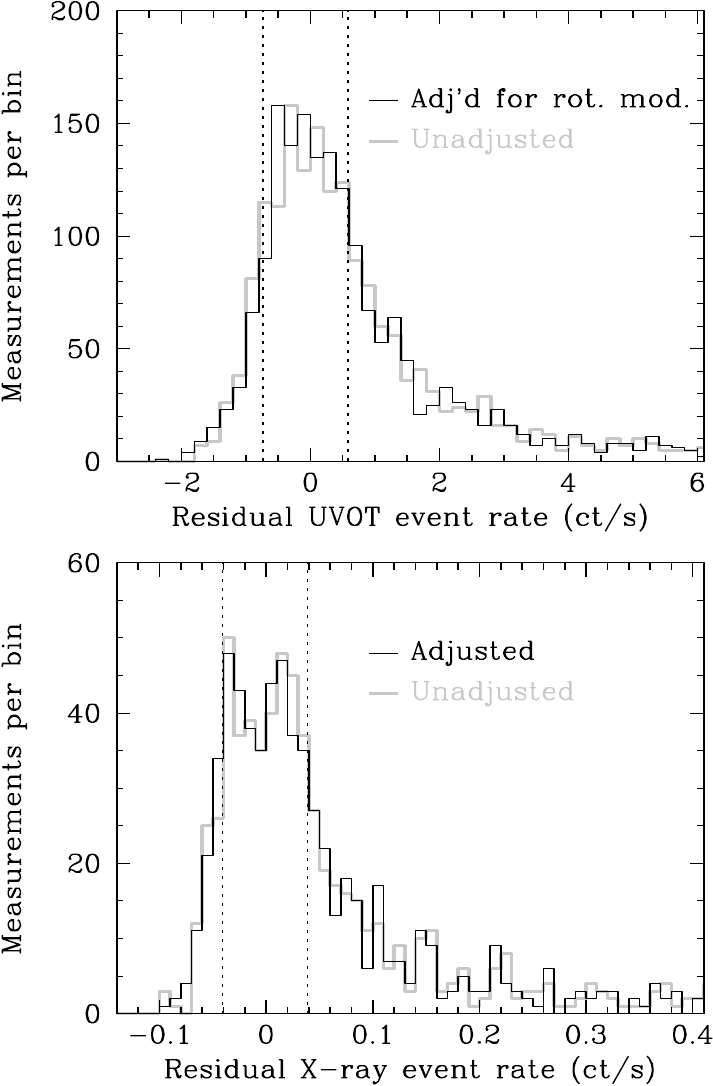}
  \end{minipage}
    \caption{Residual rates for UV/W1 (left) and XRT (center) versus
residual optical intensities for the GPR fit (excluding AOs 6 and 7). 
Upper limits on vertical axes are 95th percentiles.
%showing least-squares fits
%(solid black lines); cyan lines show $1\sigma$ uncertainties.  Fits used
%only data within the 10--60th percentiles (dotted lines) and only when
%an optical measurement was made within 21 d of a Swift measurement (black
%points).  
Solid black lines show least-squares fits to data within the 
the 10--60th percentiles (bounded by dotted lines) 
and only when an optical measurement was made within 21 d of a Swift 
measurement (black points).  
Cyan lines show $1\sigma$ uncertainties on the fits.
Right panels show binned rates using only the black points, with and
without adjustments for rotational modulation derived from the fits.
%Note that the distributions without rotational adjustments (gray) 
%are less peaked than the adjusted curves.
Dotted vertical lines mark the 10th and 60th rate percentiles after adjustments
for rotational modulation.
\label{fig:SwiftOptCorrelate}}
\end{figure*}

With a continuous model of the optical light curve, we can
now compare optical brightness with the X-ray and UV intensities
measured by Swift.
To isolate rotational modulation from longer-term, especially cyclical, 
intensity variations we subtract epoch-average rates from UV and X-ray 
measurements, and a sliding 84 day boxcar average from optical magnitudes.
``Residual'' intensities are plotted in 
Figure~\ref{fig:SwiftOptCorrelate}.
Colors distinguish between cases where there was an optical measurement
within 21 days (1/4 rotation period) of a Swift observation
so the GPR fit should be reliable,
42 days (less reliable, but only AO 13 had a significant number of such cases), 
or longer (not reliable; AOs 5 and 8 were dominated by these).  
%Points for Swift observations
%that have optical observations within 21 days ($\sim$ one-quarter of a 
%rotation period), so that the estimated magnitude from the GPR fit should be
%pretty reliable, are marked in black.  Points with up to 42 d correspondences
%(less reliable) are green, and those with longer observational mismatches 
%are orange.  

%%%%%%%%%%%%%%%%%%%%
%%% Figure -- UVOT vs optical, and maybe X-ray vs optical
%%%%%%%%%%%%%%%%%%%%%%%%%%%%%%%

To fit the optical/Swift correlations we used 10--60th
percentile rates of the Swift data to represent its quiescent emission
and varied the slope of a line through the origin, minimizing the least
squares difference between the line and the quiescent points.
Only points within a 21 d optical-Swift interval were fit,
to minimize the introduction of errors from the optical GPR curve.
%Essentially, the fitting process amounts to least-squares
%minimization while subtracting the sloped line from 
%the Swift measurements and adjusting the sample of fit points
%as rates move into or out of the 10--60th percentile range.

Results are shown in Figure~\ref{fig:SwiftOptCorrelate}.
The optical/UV correlation is quite strong, with the fit yielding
\begin{equation}
\Delta R_{W1} = (12.40 \pm 0.61)\Delta g.
\end{equation}
%where $\Delta R_{W1}$ and $\Delta g$ are the axes in Figure ***.
The true significance of the correlation, however, is somewhat less because
we had to estimate the optical brightness at the time of each Swift
measurement using the GPR fit.  Secondly,
the measurements are not entirely independent---flares
take some time to decay---and the UV rates do not follow a normal
distribution since flares skew it to higher rates and we only sample
the ``quiescent'' core.  As seen in Figure~\ref{fig:SwiftOptCorrelate}
(upper right panel),
however, the distribution is close to normal when ignoring
the high-rate tail, as we do.  If we mirror the low-rate side 
(0--10th percentile) to the high side ($>60$th percentile) 
and use that pseudo-Gaussian full distribution (as opposed
to just the core) to calculate
errors we obtain an uncertainty of 0.97 in the slope, for
a formal significance of $12.8\sigma$.
%% ** Of course that ignores errors in the GPR fit. 
%% No idea how that could be practically accounted for.

We can then use the anti-correlation relationship
between optical intensity and UV rates 
to adjust the latter for rotational modulation and reveal cleaner
measurements of long-term intensity variations arising from the
stellar cycle.  As hoped for given our efforts to sample Proxima's
emission evenly from all sides and balance any rotationally induced
excess or deficit emission, net effects on quiescent UV rates for each 
epoch are minor, ranging from -1.17\% (AO 17) to +0.85\% (AO 15D).  
Adjustments were
applied to all Swift UV measurements in AO 13, including those with
optical-Swift gaps of up to one-half rotation period (42 d);
rates for AOs 5 and 8 were not modified because of the lack of 
sufficiently contemporaneous optical measurements.

% 
% I then determined the 10th and 60th percentile rate limits for the
% set, just like I'd done to define the quiescent emission for each
% epoch separately.  You can see in findslopeBINNED.pdf that this
% picks out the core of the "quiescent" rate distribution, as intended.
% I then varied the slope of a line going through the origin, also varying
% which points were included in the 10-60th percentiles because of
% the slope, and minimized least squares differences of those points
% from the sloped line to obtain the best fit line you see.  Dotted lines
% mark the 10 and 60th percentiles, and cyan shows 1 sigma limits
% on the slope.
% 
% It's about a 12 sigma result for UV data, so I used the optical/UV
% correlation to adjust the UV rates, not including AOs 5 or 8 since
% they didn't have optical data that were close enough to Swift's obs.
% The effect ranged from -1.24% to +0.43% vs the cycle amplitude
% of roughly 20%. I did the same analysis for the X-ray data but the correlation
% was less than 2 sigma, amounting to no more than a 1.5% adjustment
% vs a cycle amplitude of ~50%, so I didn't do any adjustments.
% 

The same analysis was applied to optical/X-ray data, finding weaker
anti-correlation.  The fit slope was 
% old value before doing stellar contamination correction = 0.150
0.346, and error calculations
employing an intensity distribution with mirrored 0--10 percentiles
%employing a pseudo-Gaussian intensity distribution 
(as was done for
the UV data) indicate an uncertainty of 
% old results:  0.087, or a $1.7\sigma$ result.
0.095, or a $3.6\sigma$ result.
Net effects on average quiescent fluxes are somewhat larger
than in the W1 band but still small, ranging from
%-2.9\% (AO 17) to +4.1\% (AO 15D).
-2.9\% to +4.1\% (see Table~\ref{tab:UVOTaveRates}).
One would expect that W1 and X-ray adjustments should roughly
track each other, but even when ignoring Epoch 15 (because of
the difference in W1 and X-ray sampling; see Table~\ref{tab:SwiftObss}),
correlation is poor.  
We also found that using a slightly different, ``peakier,'' but similarly
justifiable GPR function yielded only a $1.6\sigma$ optical/X-ray correlation.
(Negligible difference in optical/UV correlation was seen using that function.)
Because of the relatively low and uncertain significance of 
the X-ray/optical correlation we consider any adjustments
of questionable value and use uncorrected X-ray intensities in subsequent 
analysis.

A possible explanation for the relatively weak correlation of X-ray emission
with optical/UV is that X-rays come from the optically thin and
spatially extended corona whereas
optical (photospheric) and UV emission (chromospheric) are essentially 
surface phenomena and therefore more subject to rotational modulation.
X-ray modulation only occurs for the fraction of emission that 
originates close to the stellar surface, 
and would be especially weak in systems where the
corona is relatively large compared to the stellar diameter (as in
M stars), and also those with smaller
inclinations (more ``pole on''); a recent paper 
\citep{Klein2021} reports Proxima's inclination as $47\pm7$ degrees.
\section{Results}
\label{sec:Results}
%%%%%%%%%%%%%%%%%%%%%%%%%%%%%%%%%%%%%%%%%%%%%%%%%%%%%%%%%%%%%%%
%%%%%%%%%%%%%%%%%%%%%%%%%%%%%%%%%%%%%%%%%%%%%%%%%%%%%%%%%%%%%%%
\subsection{Lomb-Scargle Period Analysis}
\label{sec:ResultsLSperiod}
%%%%%%%%%%%%%%%%%%%%%%%%%%%%%%%%%%%%%%%%%%%%%%%%%%%%%%%%%%%%%%%

%3.1 Lomb-Scargle Period analysis
%3.2 Stellar Cycle Period
%3.3 Cycle Amplitudes and Rossby Number
%3.4 CMEs

%%%%%%%%%%%%%%%%%%%%
%%% Figure -- Optical L-S periodogram
\begin{figure}
    \centering
    \includegraphics[width=1.00\linewidth]{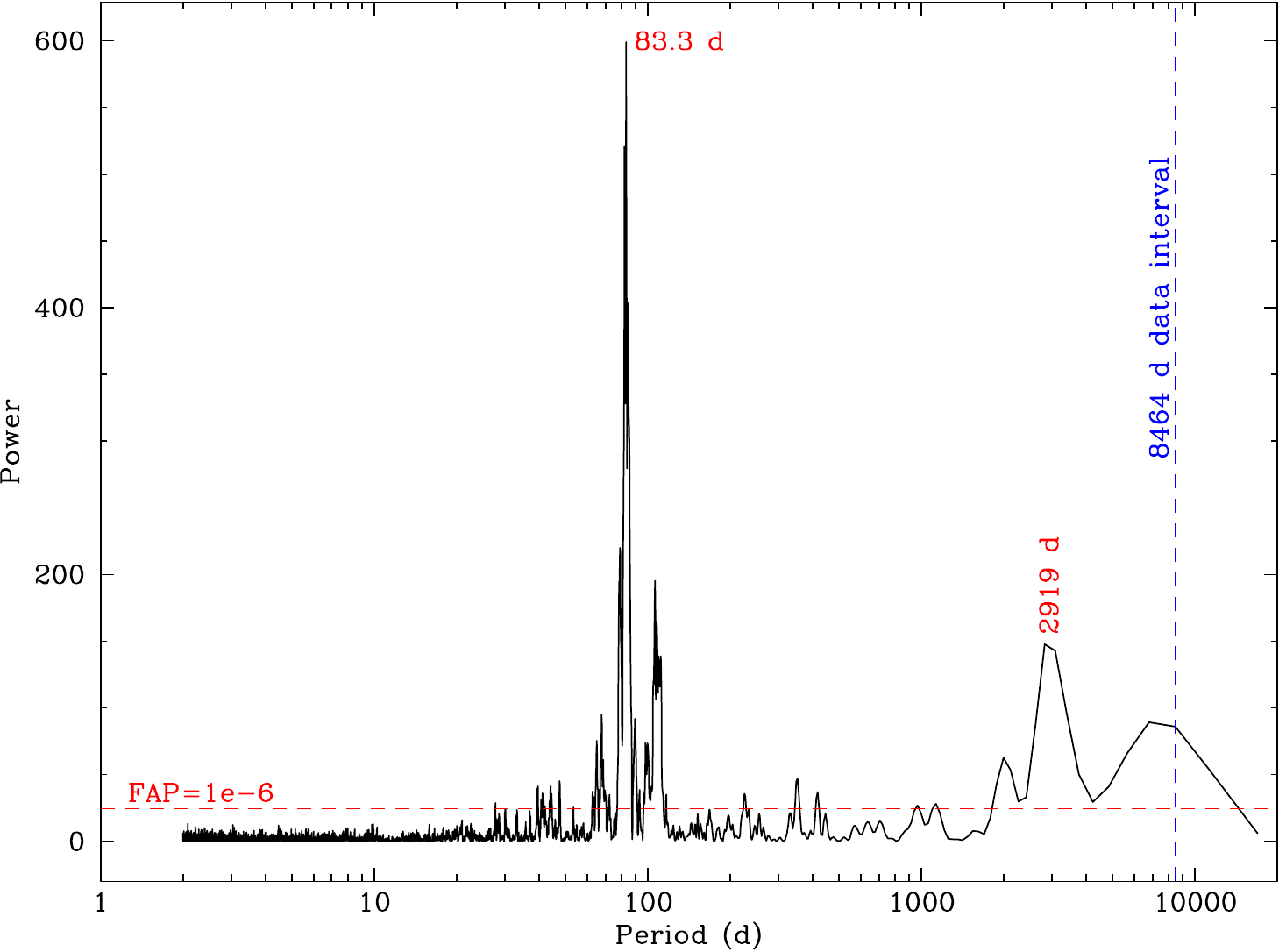}
    \caption{Floating-mean Lomb-Scargle periodogram of the full optical data 
set showing periodogram power vs. period.  The strongest rotational peak 
(83.3 d) and the cycle peak (2919 d) are labeled, as is the data timespan 
(8464 d).  
%Several other peaks clustered near the tallest one appear valid 
%(i.e., not beat periods, harmonics or window aliases) and thus may indicate 
%several rotation periods are present, i.e., evidence for differential rotation.
    \label{fig:LSperiodogram}}
\end{figure}
%%%%%%%%%%%%%%%%%%%%%%%%%%%%%%%
%%%%%%%%%%%%%%%%%%%%

%%%%%%%%%%%%%%%%%%%%
%%% Figure -- Cycle fits for X-ray and UVOT
\begin{figure}
    \centering
    \includegraphics[width=1.00\linewidth]{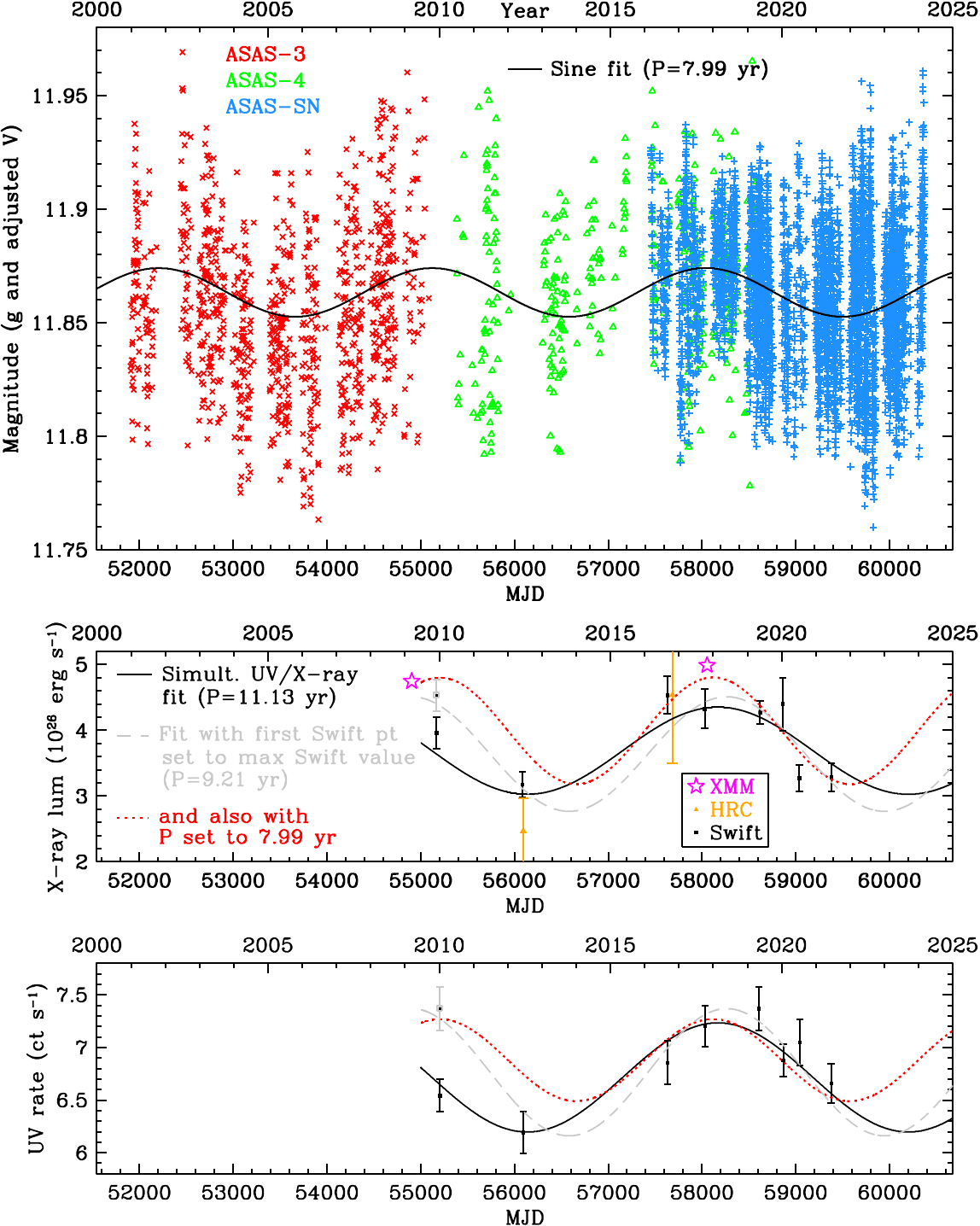}
    \caption{
	Cycle fits for optical, X-ray, and UV data.
	Optical brightness scale is inverted to aid comparison
	with the anti-correlated X-ray and UV intensities.
	X-ray and UV data were fitted simultaneously, with uncertainties
	adjusted to provide equal statistical weight.
	UV/X-ray period agreement with the optical data is much improved if
	the first ($\sim$2010) data points are raised to match peak values
	near the $\sim$2018 maximum (see text). XMM and Chandra HRC
	data points are not used for fitting.
    \label{fig:SwiftCycleFit}}
\end{figure}
%%%%%%%%%%%%%%%%%%%%%%%%%%%%%%%
%%%%%%%%%%%%%%%%%%%%

With the final adjusted values for the optical, UV, and X-ray measurements 
in hand, we can examine Proxima's long term behavior and compare it with
previous reports of a 7 yr stellar cycle. We employed a floating-mean Lomb-Scargle
periodogram for the analysis \citep[][]{Scargle1982} with 
False Alarm Probabilities (FAPs) computed following 
\citet{Scargle1982} and \citet{Horne1986}.  
%\textcolor{red}{(
%Steve edit/write about Lomb-Scargle analysis of the optical data--
%what I've written here is just to give me some scaffolding
%for other text to refer to.)
%A modified Lomb-Scargle periodogram analysis (**REF**;
The result (Figure~\ref{fig:LSperiodogram}) shows a dominant
peak at 83.3 d corresponding to the rotation period, with multiple
nearby peaks and perhaps parts of the cluster around 110 d likely deriving 
from  various 
manifestations of differential rotation and starspot movement, 
formation, and dissipation. 
Some peaks, though, we identify as beat effects between rotation 
and the year and half year observing windows and the cycle period. 
Other peaks are from the window function, harmonics, and aliases, 
except for the peak at 2919 d (7.99 yr) which
we take to be the stellar cycle period.
The broad bump at several thousand days is likely caused by the
$\sim$23 yr data interval.  
%If we (naively!) assume all periods near 83.3 d
%that are not aliases or beat periods to be due to differential rotation, 
%we find a fractional differential rotation of $(P_{\rm rot,max} - P_{\rm rot,min})/83.3 \approx 0.40$ or $3\sigma_{P_{\rm rot}}/83.3 \approx 0.35$, 
%a notably large value. If the cluster of peaks around 110 d is due to rotation,
%the gap between there and the larger cluster of periods around 83 d
%suggests a second, separate latitude band.

\vspace{10mm}
%%%%%%%%%%%%%%%%%%%%%%%%%%%%%%%%%%%%%%%%%%%%%%%%%%%%%%%%%%%%%%%
\subsection{Stellar Cycle Period and Amplitudes}
\label{sec:SineCycle}
%%%%%%%%%%%%%%%%%%%%%%%%%%%%%%%%%%%%%%%%%%%%%%%%%%%%%%%%%%%%%%%

The X-ray and UV data sets have too few points for useful L-S analysis, but they
and the optical data are suitable for sine fits to the multi-year cycle
(provided that sinusoidal
behavior is sufficiently stable over multiple cycles; see below).
ASAS-3 data yield a period of $7.34\pm0.37$ yr with an amplitude of 0.019
magnitudes (3.8\% peak-to-peak cycle variation); adding the relatively
sparse ASAS-4 data increases the period to 10.7 yr but with a very large
increase in uncertainty ($\pm4.7$ yr).  ASAS-SN data
%, with or without adding ASAS-4, yield $P_{rot}=7.9$ yr 
yield $P_{rot}=7.88$ yr 
and half the amplitude found for ASAS-3, 
while the full data set, spanning three cycles, 
yields $P_{rot}=7.99\pm0.17$ yr (the same as the L-S periodogram result) 
and an amplitude of $0.0107\pm0.0006$ mag
(Figure~\ref{fig:SwiftCycleFit}).

%Simult UV/X-ray fits
%                c0              LXmax/min       UVmax/min UVamp/c0
%Unaltered data  2015.35         1.438           1.167   0.0772
%Raised AO5      2016.09         1.63            1.196   0.0892
%Fixed P=7.99yr  2016.00         1.546           1.120   0.0567

As seen in the middle panel of that figure,
the HRC X-ray results are consistent with Swift results but their
uncertainties are quite large, even though the error bars shown
are only statistical and do not include systematic errors
(primarily from sampling).
The two XMM points are also fairly consistent with Swift, particularly 
when considering that they are expected to be a bit higher than the
Swift points (see discussion at end of Section~\ref{subsubsec:DataXMM}),
and their systematic errors are likely larger 
than those for Swift.
Because of the general consistency of results from XMM, HRC, and Swift,
and the larger, less well understood uncertainties on the XMM and HRC points,
we only use the Swift data in our cycle fits.

Independent fits to the Swift UV and X-ray data gave 
periods of 11.84 and 9.74 yr,
respectively.  A simultaneous fit to both sets of data gave poor results 
for the UV, with a visibly suppressed amplitude.  To apply equal weights
to both wavebands we therefore scaled the uncertainties to yield reduced
$\chi^2$ of 1 in separate fits, which required UV errors to be a factor of 
1.17 smaller than listed in Table~\ref{tab:UVOTaveRates} and X-ray 
errors to be 1.25 times larger than in Table~\ref{tab:XrayFits}.
The combined fit then gave $P_{rot}=11.13\pm0.53$ yr,
with $L_X^{max}/L_X^{min}=1.44$, and
%1.167 for W1, i.e., 
a cycle amplitude of 7.7\% (max/min=1.167) for the W1 band
(see black curves in bottom panels of Figure~\ref{fig:SwiftCycleFit}).

The fitted UV/X-ray cycle period of 11 yr is longer than the optical
8 yr period, but that from UV/X-ray data is much more uncertain, mostly
because of the fit's sensitivity to the first point in each band, which
is because of the limited span of the Swift measurements and data sparseness
prior to 2016.
Based on the more reliable optical 8 yr period, the UV/X-ray cycle should
be near its maximum in 2010.
We suggest that the high energy AO 5 intensities are lower than 
they ``should be'' either
because of statistical fluctuations or more likely because the cycle amplitude 
at that time is truly lower than at later times; such cycle variability is
common for the Sun and other stars.  
We also note from the previously mentioned separate fits to 
ASAS and ASAS-SN data
that Proxima's optical cycle amplitude was twice as large for earlier times
(prior to c.~2010) as later (2017 and after),
plausibly consistent with (opposite sense) cycle amplitude
variations at UV/X-ray energies.

%Simult UV/X-ray fits
%                c0              LXmax/min       UVmax/min UVamp/c0
%Unaltered data  2015.35         1.438           1.167   0.0772
%Raised AO5      2016.09         1.63            1.196   0.0892
%Fixed P=7.99yr  2016.00         1.546           1.120   0.0567

If one therefore assumes that the AO 5 measurements around 2010 were indeed 
made near a cycle maximum and sets the UV and X-ray intensities to match
those around the 2018 cycle maximum to accommodate the implicit assumption
of constant amplitude in sinusoidal fitting, the simultaneous UV/X-ray fit 
then yields $L_X^{max}/L_X^{min}=1.63$, W1 amplitude of 8.9\%,
and a significantly shorter period
of $9.21\pm0.33$ yr (dashed gray curves in Figure~\ref{fig:SwiftCycleFit}).
%and better agreement with the optical cycle.
A fit using the altered AO 5 values and freezing the period at the optical 
fit's 7.99 yr reduces $L_X^{max}/L_X^{min}$ to 1.55 and the W1 amplitude
to 5.7\% (dotted red curve).
Figure~\ref{fig:SwiftCycleFit} also shows that the optical and
UV/X-ray cycles are opposite in phase, as one would expect given the
anticorrelation seen in rotational modulation.
%27\% 
%% These next lines are for fit with P=7.99 and unaltered AO5 values:
%but the UV amplitude by a lot (82\%), mostly because
%the measured UV intensity variations seem to lag the optical 
%(and X-ray) cycle by several months, although there are not enough data 
%to tell if this is really the case.

In summary, from the 23 yrs of optical data, we find a well constrained
cycle period of 8.0 yr,
with an average amplitude of 0.011 magnitudes (0.022 mag peak-to-peak)
that has declined from 0.019 mag in the 2000's.  
The cycle is less well constrained in the W1 and X-ray bands
but consistent with an 8 yr period.
Quiescent X-ray luminosity averages 
$3.7\times10^{26}$ erg s$^{-1}$ with $L_X^{max}/L_X^{min}\sim1.5$ over the
past decade, and perhaps a little less prior to c.~2013.  
Cycle amplitude in the UV W1 band is $\sim$8\% ($\sim$17\% peak-to-peak),
and like X-ray emission, was likely weaker in the prior cycle.

%%%%%%%%%%%%%%%%%%%%%%%%%%%%%%%%%%%%%%%%%%%%%%%%%%%%%%%%%%%%%%%
\subsection{X-Ray Cycle Amplitudes and Rossby Number}
\label{sec:AmpVsRo}
%%%%%%%%%%%%%%%%%%%%%%%%%%%%%%%%%%%%%%%%%%%%%%%%%%%%%%%%%%%%%%%

Although only seven stars had measured X-ray cycles at the time,
\citet{Wargelin2017} noted that cycle amplitudes were approximately 
proportional to Rossby number ($Ro = P_{rot}/\tau_C$,
where $\tau_C$ is the timescale for convection), and that this
correlation held even for fully convective Proxima.
This relationship is similar to the finding by \citet{Wright2016} that the 
$L_X/L_{bol}\propto Ro^{-2.7}$ rotation-activity relationship for
partially convective stars below the saturation regime
\citep{Wright2011} also
applies to Proxima and three other fully convective stars.
Some of the seven stars now have updated amplitudes,
%When compared with other X-ray cycle amplitudes, Proxima follows the general 
%trend found in \citet{Wargelin2017} of increasing amplitude with larger
%Rossby number ($Ro = P_{rot}/\tau_C$,
%where $\tau_C$ is the timescale for convection).
and an additional three stars have had their 
X-ray cycles measured since then (see Table~\ref{tab:XrayCycleStar}): 
$\tau$ Boo \citep{Mittag2017}, 
$\epsilon$ Eri \citep{Coffaro2020}, 
and the rapid rotator, AB Dor \citep[][]{Singh2024}. 

A number of F stars such as $\tau$ Boo have recently been discovered from 
\caii\ measurements to have very short cycles of under one year.  
\citet{Mittag2019} found cycles ranging from 180 to 309 d in three or 
four stars, and $\tau$ Boo has an even shorter cycle of $\sim$120 d 
\citep{Mittag2017}. There are also more than 30 XMM observations of 
$\tau$ Boo, with at least one per year from 2000 to 2011.  X-ray 
data are unfortunately not dense enough to provide reliable independent 
measurements of cycle periodicity, but the range of intensities does 
provide a good indication of cycle amplitude.
Photometric ($V$-band) data likewise provide a high-confidence measurement of 
cycle period for AB Dor \citep{Singh2024}, but X-ray data (primarily from XMM)
are again too sparse to support confident period analysis, while abundant
enough to provide an upper limit on X-ray cycle amplitude.

%When X-ray cycle amplitude is plotted against Rossby number 
%$P_{\rm rot}/\tau_{\rm C}$ (Figure~\ref{fig:X-rayAmpVsRo}),
%we find Proxima near the transition between minimum amplitudes at low Ro 
%and the regime where amplitudes increase with Ro (less active stars). 

%In the Sun, long cycles tend to be weaker (or is it the opposite?).
%Does that hold for other stars? Does it depend on dynamo type,
%active vs inactive branch, etc.?
% Not enough data to say.

%%%%%%%%%%%%%%%%%%%%
% table* uses the 2-column version of table
\begin{deluxetable*}{lcccccccl}[ht!]
\tablecaption{Stars with published X-ray cycles.
\label{tab:XrayCycleStar}
}
\tablewidth{0pt}
\tablehead{
\colhead{Star Name}     
                &\colhead{Type}  
                     &\colhead{$B-V$}
                        &\colhead{$P_{cyc}$}
                                &\colhead{$P_{rot}$}
                                        &\colhead{$\tau_C$\tablenotemark{a}}
                                                &\colhead{$Ro=$}   
                                                        &\colhead{$L^{max}_{X}/$}   
                                                                & \colhead{References}   \\
%\vspace{-6.5mm} \\                                                              
\colhead{}      &\colhead{} 
                     &\colhead{($V-K_s$)}
                        &\colhead{(yr)}    
                                &\colhead{(d)}      
                                        & \colhead{(d)}       
                                                &\colhead{$P_{rot}/\tau_{C}$}   
                                                        &\colhead{$L^{min}_{X}$}   
                                                                &\colhead{}        \\
%\vspace{-9.5mm} \\                                                 
}
\startdata
% tau Boo B-V=0.48 --> Gunn table says 4.5 d for tau_C
$\tau$ Boo      &F6V & 0.508 & 0.33, 11.6\tablenotemark{b}   & 3.05  & 7.6\tablenotemark{c} & 0.40  & 1.77 & \cite{Mittag2017}  \\
$\iota$ Hor     &F8V    & 0.561 & 1.6    & 8.2   & 11.8, 23.5   & 0.695, 0.349  & 1.9 & \cite{Sanz-Forcada2019} \\
Sun             &G2V  & 0.653   &11     & 25.4  & 17.2  & 1.48  & 3.91 & \cite{Ayres2020} \\
HD 81809        &G1.5IV-V  & 0.642  &8.2    & 40.2  & 22.8\tablenotemark{d} & 1.76        & 4.9 & \cite{Orlando2017} \\
$\alpha$ Cen A  &G2V  & 0.697  &19     & 28    & 19.9  & 1.41  & 2.7 & \cite{Ayres2020}  \\
AB Dor          &K0V  & 0.830 &19.2	& 0.51	& 26.4, 53.5	& 0.0193, 0.0095	&$\la$1.4\tablenotemark{e} & \cite{Singh2024} \\
$\alpha$ Cen B  &K1V  & 0.902  &8.4    & 37    & 29.8  & 1.24  & 3.5 & \cite{Ayres2020}  \\
$\epsilon$ Eri  &K2V  & 0.881  &2.9    & 11.1  & 29.1, 57.9  & 0.38, 0.19  & 1.5 & \cite{Coffaro2020} \\
61 Cyg A        &K5V  & 1.158  &7.3    & 35.4  & 38.0  & 0.93  & 3.0 & \cite{Robrade2012}  \\
Proxima         &M5.5V &  (6.75) &8.0   & 84    & 326.2    & 0.26  & 1.5 & This work  \\
\enddata
\tablecomments{
\vspace{-1.0mm}
\tablenotetext{a}{Following the method of \citet{Irving2023}; see text for details. Second value (where present) is full 
%\textcolor{red}{(Is this supposed to be "for"?)}   No, its full CZ, not just part (the tachocline boundary region)
convective zone $\tau_{\rm C}$ (except Proxima,  where it is the only value), first value is the local (tachocline) $\tau_{\rm C}$. }
%based on B-V, with extension to M dwarfs following \citet{Gilliland1986}.}
\vspace{-1.0mm}
\tablenotetext{b}{From \citet{Baliunas1995}.}
\vspace{-1.0mm}
\tablenotetext{c}{$\tau$ Boo has a close, massive, hot Jupiter; $P_{\rm rot}$ is synchronized to $P_{\rm orbit}$, possibly affecting activity.}
%is for a main sequence star, which may not be appropriate. 
\vspace{-1.0mm}
%\tablenotetext{d}{\citet{Egeland2018} argues that the cycling
%component of this binary is evolved, our scaling of $\tau_{\rm C}$ to account for evolution make it less certain; see text.}
%% Steve, I changed the above line to the line below--is it as intended?
\tablenotetext{d}{\citet{Egeland2018} argues that the cycling
component of this binary is evolved; our scaling to account for the star's evolution makes its $\tau_{\rm C}$ value less certain. See text.}
\vspace{-1.0mm}
\tablenotetext{e}{\citet{Singh2024} find multiple superposed cycles in AB Dor
which prohibits a firm determination of the dominant cycle's amplitude.}
%$^d$Gaidos et al.~(2000).\\
%$^e$Suarez-Mascareno et al.~(2016).
%$^f$Clements et al.~(2017).
%$^g$Kiraga \& Stepie\'{n} (2007).
%$^h$Lehtinen et al.~(2016).\\
%{\bf\textcolor{red}{*** Are we using consistent $\tau$'s for all these? Do we need to discuss our selection of $\tau$ calcs/refs in the text?}}
}
\end{deluxetable*}
%%%%%%%%%%%%%%%%%%%%%%%%%%%%%%%%%

In additional to adding new stars,
we have also updated all stars' values for $\tau_{\rm C}$,
and include consideration
of recent suggestions that dynamos in active, partially convective stars
may not be solely driven by the ``local'' tachocline
dynamo but also, or perhaps even {\it instead}, by a ``global'' full convection zone dynamo; see
discussion in \citet{Irving2023} and references therein.
To determine $\tau_{\rm C}$ for all those stars
we follow the procedure of \citet{Irving2023}, and
estimate $T_{\rm eff}$ from $V - K_S$ (M dwarfs) or $B- V$ (all
others) from the tables in \citet{Pecaut2013}.  These $T_{\rm eff}$ were then
used to interpolate local $\tau_{\rm C}$ and
global $\tau_{\rm C}$ from \citet{Landin2023}.
For the lowest mass stars (nearly fully and fully convective objects;
here, only Proxima), because of difficulties in detailed interior models
\citep[see ][for discussion]{Irving2023}, we prefer the scaling relations of
\citet{Corsaro2021}, which we scaled to match the  $\tau_{\rm C}$ of 
\citet{Landin2023} at their juncture.  Thus, for these stars 
$\tau_{\rm C} = 2.9\times10^{7}
(M/M_\odot)^{1/3} (T_{\rm eff}/T_{\rm eff,\odot})^{-4/3}$ days,
and the mass was similarly interpolated from \citet{Pecaut2013} using the
estimated $T_{\rm eff}$.  

%\citet{Egeland2018} found HD 81809 is a binary whose
%emission is dominated by a subgiant. We therefore modified its
%$\tau_{\rm C}$ to be appropriate for its evolved state by determining 
%the factor $\tau_{\rm C}$ increased relative to the main sequence
%for a 1.6$M_\odot$ star with its current $T_{\rm eff}$ in \citet{Gunn1998} 
%(their Fig. 3).  This factor increase was then applied to the \citet{Landin2023}  model.

\citet{Egeland2018} found that HD 81809 is a binary whose
emission is dominated by a subgiant. We determined a
$\tau_{\rm C}$ appropriate for its evolved state by
taking the \citet{Landin2023} value for a 1.6$M_\odot$ main sequence star
and then applying a scaling factor equal to the ratio between
the values listed by \citet{Gunn1998}
for a star with HD 81809's current $T_{\rm eff}$
and for its main sequence counterpart (their Figure 3).

% Steve's original text--altered above on 6/20/24
%To determine $\tau_{\rm C}$ for all those stars
%we follow the procedure of \citet{Irving2023}, and
%estimate $T_{\rm eff}$ from $V - K_S$ (M dwarfs) or $B- V$ (all
%others) from the tables in \citet{Pecaut2013}.  These $T_{\rm eff}$ were then
%used to interpolate local (tachocline dynamo) $\tau_{\rm C}$ and
%global (full convection zone dynamo) $\tau_{\rm C}$ from \citet{Landin2023}.
%For the lowest mass stars (nearly fully and fully convective objects;
%here, Proxima), because of difficulties in detailed interior models
%\citep[see ][for discussion]{Irving2023}, we prefer the scaling relations of
%\citet{Corsaro2021}, which we scaled to match the  $\tau_{\rm C}$ of 
%\citet{Landin2023} at their juncture.  Thus, for these stars 
%$\tau_{\rm C} = 2.9\times10^{7}
%(M/M_\odot)^{1/3} (T_{\rm eff}/T_{\rm eff,\odot})^{-4/3}$ days.
%Here, the mass was similarly interpolated from \citet{Pecaut2013} using the
%estimated $T_{\rm eff}$.  \citet{Egeland2018} found HD 81809 is binary whose
%emission is dominated by a subgiant. We therefore modified its
%$\tau_{\rm C}$ to be appropriate for its evolved state by determining 
%the factor $\tau_{\rm C}$ increased relative to the main sequence
%for a 1.6$M_\odot$ star with its current $T_{\rm eff}$ in \citet{Gunn1998} 
%(their Fig. 3).  This factor increase was then applied to the \citet{Landin2023}  model.

\begin{figure}[ht]
    \centering
    \includegraphics[width=0.98\linewidth]{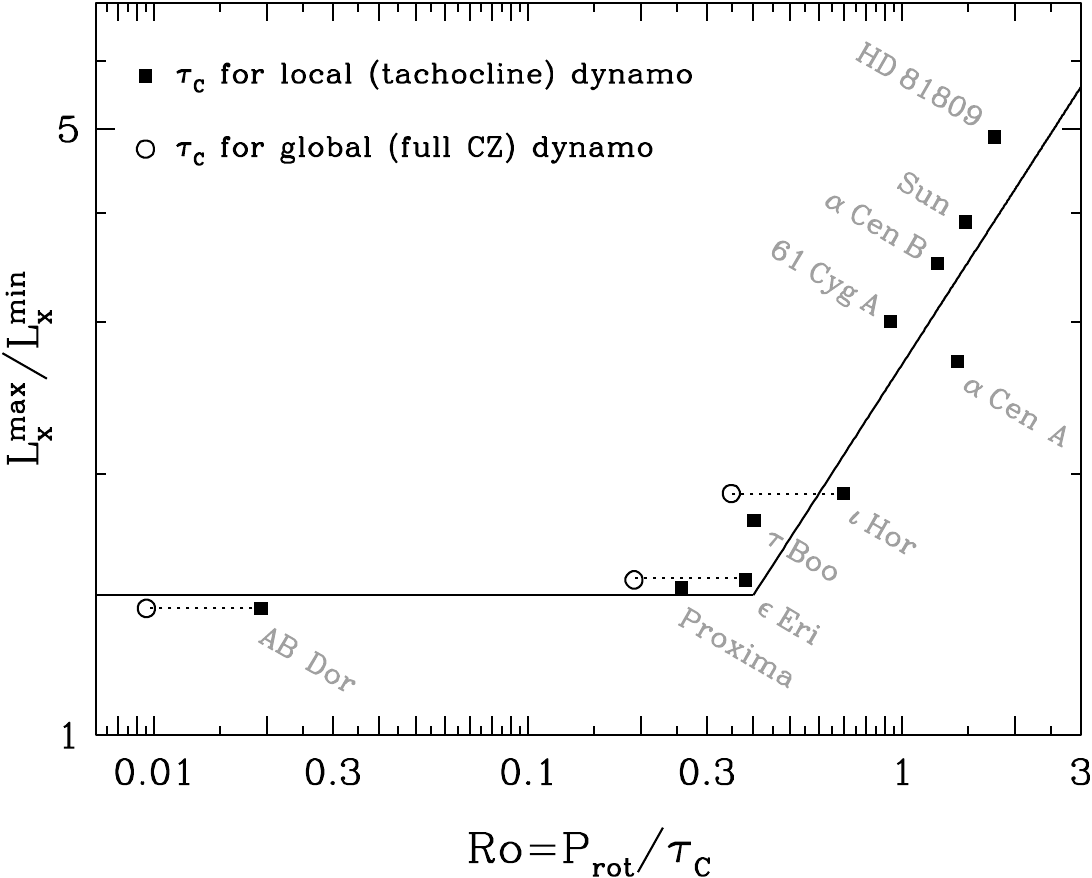}
    \caption{
        X-ray cycle amplitude vs Rossby number, using values listed in
Table~\ref{tab:XrayCycleStar}. 
Three stars have pairs of points, reflecting uncertainty in whether 
the local (tachocline) or global (full convective zone) dynamo is dominant 
\citep[see ][]{Irving2023}.  
Amplitudes for Proxima and $\epsilon$ Eri
have been slightly shifted for clarity.
    \label{fig:X-rayAmpVsRo}}
\end{figure}

Figure~\ref{fig:X-rayAmpVsRo} shows the updated results for X-ray
cycle amplitude vs Rossby number.  The same trend is found as before
but with the additional stars there is now a clear transition to a
limiting amplitude at small Ro (as must occur since amplitudes cannot
go below 1).  Proxima lies near the transition between low Ro/active
and higher Ro/less active stars.  The shape of the curve is again
reminiscent of the rotation-activity relationship, which saturates
around $L_X/L_{bol}\sim10^{-3}$ at small Ro; with the $\tau_{\rm C}$ scale used here, total X-ray emission saturates at Ro $\approx 0.11$.  Thus it would appear that cycle amplitude saturates at somewhat slower rotation: at Ro $\approx 0.4$ if active stars are dominated by a tachocline dynamo (and $\tau_{\rm C,L}$ is appropriate, as sketched in Figure~\ref{fig:X-rayAmpVsRo}) or Ro $\approx 0.2$ if a full convection zone dynamo dominates (and $\tau_{\rm C,G}$ is better). 
%{\bf\textcolor{red}{*** Note: Brad, I eyeballed where the fits "desaturate" - might want to tweak the 0.2 and 0.4.}}

%%%%%%%%%%%%%%%%%%%%
%%% Figure -- X-ray cycle  amplitude vs Rossby number
%%%%%%%%%%%%%%%%%%%%%%%%%%%%%%%

As suggested by
\citet{Wargelin2017}, the correlation of high Ro with high amplitude
(now with saturation at low Ro) can be plausibly explained by
more active stars (smaller Ro) having a greater covering fraction of
X-ray-emitting active regions even at cycle minimum, 
with less room available for additional emission at cycle maximum.  
Note that in the most extreme case, maximum amplitude is achieved with 
half the star active and half quiet, which has exactly half the total flux of the maximum total flux case (full star covered).  
This scenario is also supported, for Proxima, by a lack of
I-band variation coupled with a strong trend in V-I vs V and
the star growing redder during the optically fainter part of its
cycle, implying a large, fairly even covering fraction of cool starspots
\citep{Wargelin2017}.

%%%%%%%%%%%%%%%%%%%%%%%%%%%%%%%%%%%%%%%%%%%%%%%%%%%%%%%%%%%%%%%
\subsection{Possible Coronal Mass Ejections}
\label{sec:ResultsCME}
%%%%%%%%%%%%%%%%%%%%%%%%%%%%%%%%%%%%%%%%%%%%%%%%%%%%%%%%%%%%%%%

% See documentation in Proxima/Manuscript/Paper/CME.info
%8 day gap 817e to 818a
%1.2 hr gap 1233a to 1233b

When studying light curves and the rate distributions in
Figure~\ref{fig:OrderedRateCurves} we noticed a
few instances of extremely low rates.
Focusing on the ten X-ray bins
with rates $<$0.02 ct s$^{-1}$
(out of 856 total for the XRT),
three are from snapshots where the source is within a few pixels of
dead CCD columns (so rates are suspect)
and the other bins within the same snapshot
are not particularly low,
suggesting that the low rates are instrumental artifacts.
A fourth case is from a snapshot with a single bin, with unremarkable rates 
in nearby snapshots.

The remaining six low-rate bins, however, are particularly interesting because:
\begin{itemize}
\item The source is far away from bad pixels (so rate corrections are reliable).
\vspace{-2mm}
\item They come in two groups (three from 31676018a and three from 31676033bP,
where a, b, c... distinguish snapshots within an observation,
and P and F denote pre- and flare intervals within a snapshot)
in which {\em all} bins have low rates,
indicating a real low rate and not a statistical fluke.
\vspace{-2mm}
\item The low-rate bins are followed by large
flares ($>$1 ct s$^{-1}$), which occurs for only 1.8\% of the 856 bins.
\end{itemize}

The three low-rate bins from 018a (dropping the 31676 prefix)
contain ten counts, $6.4\sigma$ below the 60 counts expected from the
AO 8 quiescent average. Their average rate is
$0.0121\pm0.0038$ ct s$^{-1}$,
followed 1.37 hr later by a decaying flare in snapshot 018b,
with an initial rate of 1.46 ct s$^{-1}$.  
The three bins in
033bP average $0.0137\pm0.0037$ ct s$^{-1}$
($10.5\sigma$ below the quiescent average) immediately followed by the remainder
of the snapshot in a single 69 s bin with a rate of 1.09 ct s$^{-1}$.

%The three low-rate bins from 018a (dropping the 31676 prefix)
%average $0.0121\pm0.0038$ ct s$^{-1}$,
%followed 1.37 hr later by a decaying flare in snapshot 018b,
%with a first bin rate of 1.46 ct s$^{-1}$.  
%The ten counts in the three
%low-rate bins are $6.4\sigma$ below the 60 counts expected from the
%AO 8 quiescent average.
%The three bins in
%033bP average $0.0137\pm0.0037$ ct s$^{-1}$
%($10.5\sigma$ below the quiescent average) immediately followed by the remainder
%of the snapshot in a single 69 s bin with a rate of 1.09 ct s$^{-1}$.

We also examined UVOT data for similar cases, with the only notable
occurrences paralleling what was seen in the XRT.
For the 018a/b pairing, six of the ten 018a bins 
(recall the typical four UVOT time bins per XRT bin)
had rates within the lowest 0.6\% of the 2486 total bins, and all eleven 018b
bins are among those with the 3\% highest rates.
The twelve 033bP bins had rates
in the 2nd--41st percentiles (average 6.364 ct s$^{-1}$),
while the sole 033bF bin (at 70 ct s$^{-1}$)
was the second highest in the entire Swift data set.

A plausible explanation for low rates is that a coronal mass ejection (CME)
expands into the stellar corona and leaves an evacuated low-emission volume,
as is seen on the Sun where CMEs can cause a reduction in
total coronal emission \citep{Harra2016}.
With this mechanism in mind,
\citet{Veronig2021} analyzed extreme UV and X-ray archival data
from other stars and found 21 candidate CMEs where flares were
followed by significant decreases in coronal emission.
The two cases we are examining here, however, feature low rates
{\em preceding} flares, either immediately or within 1.4 hr, which
is difficult to understand in the context of a CME.

Low emission immediately preceding a flare 
{\em has} been seen in the Sun in the UV,
particularly the 171 \AA\ band \citep{Mason2014},
%% The 171 Ang light curve was grabbed to show in \citep{Loyd2022}, 
but the degree of dimming was small
compared to what occurred after the main flare.
If the low rates seen in Proxima observations are indeed connected with CMEs,
it is more likely that the low rates {\em are}
preceded by flares, but that the flares occurred in the gaps
(8 d and 1.2 hr, respectively) preceding the two groups of low-rate snapshots
and were not seen.

%%% I often had trouble with large empty spaces and/or 
%%% text that ran off the bottom of the first column near the end.
%%% This additional spacing is a kludge to sort of fix that.
%%% Hmm, although it seems like the root problem might be that the
%%% Acknowledgments start a new column(?).
\vspace{30mm}

\section{Summary and discussion} \label{sec:Summary}

This study of Proxima Cen has presented an analysis of optical photometry
(23 years), Swift X-ray and UV measurements (8 epochs over a 12 year span),
and supplementary XMM and Chandra HRC X-ray measurements,
finding a clear 8.0 yr stellar cycle with average amplitude of
0.011 mag (0.022 mag peak-to-peak)
in the optical data and similar
periodicity in the Swift data.  Proxima is by far the smallest of the few
stars to have $\sim$regular X-ray monitoring over many years, and the
clarity of its cycle and consistency of results across optical,
UV, and X-ray energies strongly supports the growing evidence for
stellar cycles among fully convective stars.
Proxima's X-ray cycle is $\sim$1.5 times
brighter at maximum than at minimum.
Among the ten stars with published X-ray cycles,
a strong correlation between cycle amplitude
and Rossby number is found, with amplitude decreasing toward
smaller Ro before plateauing at a value not far above one.

In the UV W1 band centered around 2800 \AA,
cycle amplitude is $\sim$8\% (max/min=1.17).
Over a cycle, X-ray/UV intensity is anti-correlated
with optical brightness, as is also true for rotational modulation which has
very strong optical periodicity around 84 d.  We applied corrections
for contamination by other stars in the optical and UV bands as Proxima
moves across its relatively crowded field, yielding clean light curves
for the periodicity and correlation analyses.

Significant flaring is present in the UV and X-ray bands, and quiescent
emission levels were computed using 10th to 60th percentiles.  Quiescent
X-ray emission over a cycle averages $3.7\times 10^{26}$ erg s$^{-1}$,
and inspection of the Swift X-ray light curves yielded two instances of
statistically significant anomalously low emission 
%followed by large flares 
that may be associated with coronal mass ejections.

%\begin{acknowledgments}
\section*{Acknowledgments}

This work was supported by NASA's Swift Guest Investigator program
under Grants 80NSSC17K0332, 80NSSC20K1111, and 80NSSC22K0039,
and by NASA's XMM-Newton Guest Observer program under Grant 80NSSC18K0398
and the Chandra Guest Observer program under Grant DD6-17086X.
BJW and PR were also supported by NASA contract NAS8-03060
to the Chandra X-ray Center.
 SHS  gratefully acknowledges additional support from NASA XRP grant 80NSSC21K0607  and NASA EPRV grant 80NSSC21K1037.
ZAI acknowledges support from the UK Research and Innovation's Science and Technology Facilities Council grant ST/X508767/1.
% other NASA *** grant(s).
%
% Those were AOs 13,16,17.  AO12 was TOO observations so no funding.
% https://skymapper.anu.edu.au/how-to-cite/ sez
%Authors are required to include the following acknowledgement 
%paragraph in the paper:

We thank the ASAS and ASAS-SN collaborations for providing optical
photometry data, the UK Swift Science Data Centre at the 
University of Leicester for providing Swift data and processing tools,
and the Chandra X-ray Center for the CIAO and Sherpa analysis programs.
We also thank E.~Shkolnik for helpful discussions about CMEs.
The five Chandra HRC observations used in our analysis can be accessed
via \dataset[DOI:10.25574/cdc.306]{https://doi.org/10.25574/cdc.306}.
%https://doi.org/10.25574/cdc.306

Lastly, we acknowledge use of optical data from DR4 of the 
SkyMapper Southern Survey, which has been funded through ARC LIEF
 grant LE130100104 from the Australian Research Council, awarded to the 
University of Sydney, the Australian National University, 
Swinburne University of Technology, the University of Queensland, 
the University of Western Australia, the University of Melbourne, 
Curtin University of Technology, Monash University and the
Australian Astronomical Observatory. SkyMapper is owned and operated by 
the Australian National University's Research School of Astronomy and 
Astrophysics. The survey data were processed and provided by the SkyMapper 
Team at ANU. The SkyMapper node of the All-Sky Virtual Observatory (ASVO) 
is hosted at the National Computational Infrastructure (NCI). Development 
and support of the SkyMapper node of the ASVO has been funded in part by 
Astronomy Australia Limited (AAL) and the Australian Government through the 
Commonwealth's Education Investment Fund (EIF) and National Collaborative 
Research Infrastructure Strategy (NCRIS), particularly the National 
eResearch Collaboration Tools and Resources (NeCTAR) and the Australian 
National Data Service Projects (ANDS).
%\end{acknowledgments}


\vspace{5mm}
%\facility{Swift, ASAS, ASAS-SN, CXO, Skymapper}
\facilities{Swift (XRT and UVOT), ASAS, ASAS-SN, CXO (HRC-I), Skymapper, XMM (EPIC pn)}

%\section{Data Availability}
%
%Data analyzed here are publicly available via the ASAS All-Star Catalog,
%\footnote{\url{http://www.astrouw.edu.pl/asas}} 
%the ASAS-SN light curve server,
%\footnote{\url{https://asas-sn.osu.edu}} 
%ASAS-4 data on Proxima were provided by G.~Pojma{\'n}ski.
%Swift data are available on.....
%Note sure if we want to separately cite software again...
%%\software{astropy \citep{2013A&A...558A..33A,2018AJ....156..123A},  
%          Cloudy \citep{2013RMxAA..49..137F}, 
%          Source Extractor \citep{1996A&AS..117..393B}
%          }

%%%% bibtext sample631 --- NO!  It should be bibtex, not bibtext
%% bibtex sample631

\bibliography{prox}
\bibliographystyle{aasjournal}


\end{document}